\documentclass[12pt]{article}

\usepackage{amssymb,amsmath}
\usepackage{graphicx}

\newcommand{\be}{\begin{equation}}
\newcommand{\ee}{\end{equation}}

\newcommand{\dlt}{\delta}
\newcommand{\prt}{\partial}
\newcommand{\br}{{\bf r}}

\newcommand{\bp}{{\bf p}}

\newcommand{\bt}{\beta}

\newcommand{\al}{\alpha}
\newcommand{\ra}{\rightarrow}

\newcommand{\om}{\omega}
\newcommand{\Om}{\Omega}

\newcommand{\dgr}{\dagger}
\newcommand{\lbd}{\lambda}

\newcommand{\cD}{{\cal D}}

\newcommand{\cH}{{\cal H}}

\newcommand{\rgl}{\rangle}
\newcommand{\lgl}{\langle}

\begin{document}

\begin{center}

{\Large{\bf Statistics of Multiscale Fluctuations in Macromolecular Systems} \\ [5mm] 
 
Vyacheslav I. Yukalov$^{1*}$ and Elizaveta P. Yukalova$^{2}$} \\ [3mm]

{\it 
$^1$Bogolubov Laboratory of Theoretical Physics, \\
Joint Institute for Nuclear Research, Dubna 141980, Russia \\ [3mm]

$^2$Laboratory of Information Technologies, \\
Joint Institute for Nuclear Research, Dubna 141980, Russia} 

\end{center}

\vskip 2cm

\begin{abstract}

An approach is suggested for treating multiscale fluctuations in macromolecular
systems. The emphasis is on the statistical properties of such fluctuations. The
approach is illustrated by a macromolecular system with mesoscopic fluctuations
between the states of atomic orbitals. Strong-orbital and weak-orbital couplings
fluctuationally arise, being multiscale in space and time. Statistical properties 
of the system are obtained by averaging over the multiscale fluctuations. The 
existence of such multiscale fluctuations causes phase transitions between 
strong-coupling and weak-coupling states. These transitions are connected with
structure and size transformations of macromolecules. An approach for treating 
density and size multiscale fluctuations by means of classical statistical 
mechanics is also advanced.
 
\end{abstract}

\vskip 1cm

$^*$Corresponding author: V.I. Yukalov

E-mail: yukalov@theor.jinr.ru

\newpage

\section{Introduction}

Macromolecular systems are assemblies of many atoms, ranging between hundreds 
or thousands to millions and many millions of atoms. Examples are viruses, 
ribosomes, and other assemblies, as discussed in Refs.$^{1-7}$. Similar 
properties are also exhibited by artificial macromolecules, such as 
nanoclusters$^{8-11}$, quantum dots$^{12,13}$, and various ensembles of trapped 
atoms, whose description can be found in books$^{14-16}$ and review 
articles$^{17-29}$. The nanosize objects can be employed in a variety of 
applications$^{14-29}$, including such nontrivial ones as the creation of 
quantum dot biocomposite structures using genetically engineered bacteriophage 
viruses$^{30}$.

The systems of nanosizes possess peculiar properties by combining the features
that are typical of macroscopic and microscopic systems. From one side, 
nanostructures contain many molecules and, hence, can be described by means
of statistical mechanics, as applied to bulk matter. From another side, the 
finiteness of such structures can play an important role, making them 
dependent on their surrounding and influencing stronger fluctuations of 
observable quantities.  

One of the typical properties of complex macromolecular systems is the 
existence of processes involving widely separated time and length scales. 
Such systems exhibit two main types of fluctuations: fast fluctuations, 
related to atomic collisions and vibrations, and slow fluctuations, 
representing coherent motion of many atoms simultaneously. Between the short
and long timescales, there is a gap that makes possible the distinction 
of these principally different kinds of motion. The slow coherent motion of 
many atoms is characterized by a broad range of timescales, which makes the
description of such nontrivial dynamics a complicated problem. A multiscale
analysis for describing the dynamics of nanosystems, involving Smoluchowski 
and Fokker-Planck equations, has been developed by Ortoleva et al.$^{1-7}$.   
In the latter articles, one can find more information on the coherent 
multiscale motion of groups of correlated atoms. 

Phenomenological methods of treating multiscale mesoscopic fluctuations 
occurring in condensed-matter systems have been suggested by Frenkel$^{31}$,
Fisher$^{32,33}$, and Khait$^{34,35}$.

In the present paper, we also address the problem of describing complex
macromolecular systems displaying fluctuations in a wide range of space and
time scales. However, our approach is different from that of Ortoleva et al.
We aim at considering not the dynamics of multiscale fluctuations, but their 
statistics. Our approach is also different from the phenomenological methods
of other authors$^{31-35}$, as far as we aim at developing a microscopic 
theory.
   
The realistic picture, we keep in mind, is as follows. Suppose, we observe 
the behavior of a system during the time interval that is larger than the 
typical fluctuation time of slow multiscale fluctuations, so that during 
this observation time, there arises a number of these fluctuations.
This implies that the experimental measurements will produce the smoothed 
results averaged over all these fluctuations. Therefore, we do not need to 
know the detailed dynamics of the system at each moment of the experimental 
observation time, but we need to know only the smoothed picture averaged 
over all fluctuations that have happened during this time. This is what is 
called the statistics of fluctuations. In that sense, our approach is 
complimentary to that of Ortoleva et al.$^{1-7}$. 

The paper is organized as follows. In Sec. 2, we delineate the general 
approach that will be employed in the following sections. In Sec. 3, the 
model of a macromolecular system is derived, which we shall use for 
illustrating the approach. For generality, we consider a quantum model, 
keeping in mind that many biological systems to some extent display 
quantum properties$^{36-40}$. In Sec. 4, the behavior of the system order 
parameters is analyzed. Of course, the use of quantum models is not always
necessary. Therefore, in Sec. 5, we formulate the approach in the language 
of classical statistical mechanics. In Sec. 6, using the classical 
statistics, we show how multiscale density fluctuations can provoke sharp
changes of a macromolecule size. Section 7 concludes.

\section{Microscopic Versus Mesoscopic Fluctuations}

The aim of the present section is to give the ideological and mathematical 
foundation for the statistical approach of treating multiscale fluctuations.
In Sec. 2.1, we describe how the characteristic space and time scales of 
statistical systems can be evaluated and explain in what sense multiscale 
fluctuations are mesoscopic. This picture makes it clear that the snap-shots
of the system with mesoscopic fluctuations can be described by means of the
manifold indicator functions, as is shown in Sec. 2.2. The specific feature
of the mesoscopic fluctuations of being multiscale serves as a justification
for the procedure of averaging over spatial configurations of such fluctuations,
which is formulated in Sec. 2.3. This section also provides the general 
formulas that are necessary for considering the particular models of the 
following sections.

\subsection{Length and Time Scales}

Let us consider a macromolecular system consisting of many atoms or molecules.
In order to give a precise classification of different types of motion and 
the related fluctuation types, let us, first of all, discuss the 
characteristic length and time scales typical of multiatomic systems. 

Atoms, or molecules, composing the system, interact with each other, which
introduces the interaction length or {\it scattering length}, $a_s$. The 
{\it mean interatomic distance} $a$ is connected with the average atomic 
density $\rho$ by the relation $\rho a^3 = 1$. Atoms can move without 
collisions at the distance of the {\it mean free path} 
$\lambda \sim 1/\rho a_s^2$. Groups of atoms are correlated and can move 
in a coherent way, if the group linear size is not smaller than the 
{\it correlation length} $l_c$. The correlated atoms can coherently move 
in groups whose size is denoted as the {\it fluctuation length} $l_f$. The 
largest length is a linear size $L$ of the whole system. The standard 
relations between these characteristic lengths satisfy the inequalities
\be
\label{1}
 a_s < \lbd < l_c < l_f < L \;  .
\ee
In many cases, it happens that $a_s$ is of order of $\lambda$. The 
fluctuation length $l_f$ is not just a single fixed quantity, but it can 
represent a dense set of values between $l_c$ and $L$.    
 
These characteristic lengths are connected with the corresponding 
characteristic times. The interaction time $t_{int}$ defines the duration 
of atomic interactions and is given by the ratio of the scattering length 
$a_s$ versus atomic velocity $v \sim \hbar/ m a_s$, where $m$ is atomic mass. 
The local equilibration time $t_{loc}$ is defined by the ratio of the 
mean free path $\lambda \sim 1/\rho a_s^2$ over the atomic velocity $v$. And 
the correlation time $t_{cor}$ is given by the ratio of the correlation 
length $l_{cor}$ over velocity $v$. Summarizing, we have the 
{\it interaction time}
\be
\label{2}
 t_{int} = \frac{a_s}{v} = \frac{ma_s^2}{\hbar} \;  ,
\ee
{\it local-equilibration time}
\be
\label{3}
t_{loc} = \frac{\lbd}{v} = \frac{m}{\hbar\rho a_s} \; , 
\ee
and the {\it correlation time}
\be
\label{4}
 t_{cor} = \frac{l_c}{v} = \frac{ma_sl_c}{\hbar} \;  .
\ee
The time of experimental observation will be denoted by $t_{exp}$. The typical 
relation between these times is 
\be
\label{5}
 t_{int} < t_{loc} < t_{cor} < t_{exp} \;  .
\ee
Though $t_{int}$ can be of order $t_{loc}$. The coherent motion of atomic 
groups occurs during the {\it fluctuation time} $t_f$ that is between 
$t_{cor}$ and $t_{exp}$,
$$
t_{cor} < t_f < t_{exp}. 
$$
Again, $t_f$ is not a single fixed quantity, but can densely fill the whole 
given temporal interval.

Depending on the considered time interval, the system dynamics can be separated
into several stages. The shortest is the interaction stage, corresponding to the 
time of atomic interactions,
\be
\label{6}
0 < t < t_{int} \qquad ({\rm interaction\; stage} ) \; .
\ee
The second is the kinetic stage, when atoms have experienced just a few collisions, 
but have had yet not enough time to become well correlated,
\be
\label{7}
t_{int} < t < t_{loc} \qquad ({\rm kinetic\; stage} ) \; .
\ee
After the local-equilibration time, atoms become already correlated with each 
other and can form coherently moving groups of many atoms,
\be
\label{8}
 t_{loc} < t < t_{cor} \qquad ({\rm fluctuation\; stage} ) \; .
\ee
And, when the experimental observation time is sufficiently long, there exists
the quasi-equilibrium stage, when a number of coherent fluctuations occur during
the observation time, 
\be
\label{9}
t_{cor} < t < t_{exp} \qquad ({\rm quasi-equilibrium\; stage} ) \; .
\ee
The stage is termed quasi-equilibrium, since, at each moment of time, the 
system is not at equilibrium, being nonuniform and dynamically non-stationary, 
but on the longer time scale, it can be treated as equilibrium on average, 
fluctuating around a quasi-stationary state.  

Estimating the above quantities for typical multiatomic systems of condensed
matter type$^{1-7,19,41-43}$, we have $a_s \sim a \sim \lambda \sim 10^{-8}$ cm 
and $l_c \gtrsim 10^{-7}$ cm. While for the typical times, we get 
$t_{int} \sim t_{loc} \sim 10^{-14}$ s to $10^{-13}$ s, and
$t_{cor} \gtrsim 10^{-12}$ s. 

These estimates show that there exist two principally different types of motion.
At atomic length and time scales, there are fast {\it microscopic} fluctuations
related to atomic vibrations and collisions and characterized by the typical 
times $t_{int} \sim 10^{-14}$ s. And there are slow fluctuations corresponding 
to the coherent motion of correlated atomic groups, with the typical times
$t_{cor} \gtrsim 10^{-12}$ s. The slow fluctuations occur in the diapason of 
lengths, $l_f$,  and times, $t_f$, satisfying the inequalities
\be
\label{10}
 \lbd < l_c < l_f < L \; ,  \qquad t_{loc} < t_{cor} <t_f < t_{exp} \;  .
\ee
In that sense, these coherent fluctuations can be called {\it mesoscopic}.

As has been stressed by Ortoleva et al.$^{1-7}$, there are two crucial points 
distinguishing these two types of fluctuations. First, microscopic fluctuations 
of individual atoms and mesoscopic fluctuations of correlated atomic groups 
are separated by a large gap. Thus, the former are characterized by the time 
scale of $10^{-14}$ s, while the lowest scale of the latter is of order 
$10^{-12}$ s. Second, the slow mesoscopic fluctuations do not have just one time 
scale, but they can happen in a wide temporal range, starting from $10^{-12}$ s 
and up to the observation time. That is, the mesoscopic fluctuations exhibit 
{\it multiscale motion}.

\subsection{Snap-Shot of Heterogeneous State}

Let us assume that we can make a snap-shot of the system at a fixed moment 
of time. In the presence of mesoscopic fluctuations, the system is non-uniform,
being separated into several regions with different properties. These 
properties can be characterized by order parameters or order indices$^{44,45}$.
Actually, the mesoscopic collective fluctuations are nothing but the slow 
fluctuations of the order parameters$^{1-7}$. Thus, at each moment of time, the 
system can be highly nonuniform, which is termed heterogeneous state. Generally, 
the order parameters can be associated with some phases, because of which this
heterogeneous state, with nonuniform order parameters, can also be called 
heterophase state$^{41-43}$ and the system parts can be termed the phases.
  
Making a snap-shot of a heterogeneous system, one can distinguish the spatial 
regions with different order parameters. Then the whole system volume can be 
represented as the union of these spatial manifolds:
\be
\label{11}
 \mathbb{V} = \bigcup_\nu \mathbb{V}_\nu \; , \qquad 
V = \sum_\nu V_\nu \;  ,
\ee
with the related manifold measures
\be
\label{12}
 V \equiv {\rm mes} \mathbb{V} \; , \qquad
V_\nu \equiv {\rm mes} \mathbb{V}_\nu \;  ,  
\ee
where the index $\nu = 1,2, \ldots$ enumerates the phases with different 
order parameters. The separation into the subregions can be done by invoking 
the notion of the {\it equimolecular separating surface}, when the total 
number of atoms $N$ is the sum of atoms in each of the subregions,
\be
\label{13}
 N = \sum_\nu N_\nu \;  .
\ee
The definition and properties of the equimolecular separating surface have 
been introduced and described by Gibbs$^{46,47}$ and are widely employed in 
chemical and physical literature$^{48,49}$. This separating surface is defined 
in such a way that the surface density be zero, because of which the 
additivity of the atomic numbers is exact. Another convenience of employing 
this separating surface is that the volume and all extensive thermodynamic 
characteristics of the whole system are additive with respect to the volumes 
and thermodynamic characteristics of its parts$^{46-49}$, similarly to 
property (13). The equimolecular separating surface can be defined for any 
spatial region of arbitrary size$^{46-49}$.

Mathematically, the separated spatial regions can be described by the 
manifold indicator functions$^{50}$. If the phase, labeled by the index $\nu$,
occupies the volume $\mathbb{V}_\nu$, the related manifold indicator function
is defined as
\begin{eqnarray}
\label{14}
\xi_\nu(\br) \equiv \left \{
\begin{array}{ll}
1 , &  \br \in \mathbb{V}_\nu \; , \\
0 , &  \br \not\in \mathbb{V}_\nu \; .
\end{array} \right.
\end{eqnarray}
The set
\be
\label{15}
\xi \equiv \left\{ \xi_\nu(\br) : \; \nu = 1,2, \ldots ;
\; \br \in \mathbb{V} \right \} 
\ee
of all indicator functions uniquely describes the given spatial configuration.   

The system with the fixed spatial configuration of local phases should be 
described by the Gibbs local-equilibrium ensemble$^{46,47}$. For this purpose,
it is necessary to define the local single-particle energy operators 
$\hat{K}_\nu({\bf r})$ and the local interaction energy operators 
$\hat{\Phi}({\bf r}, {\bf r}^\prime)$. The Hamiltonian of the $\nu$-phase
is written in the form
\be
\label{16}
H_\nu(\xi) = \int \left [ \hat K_\nu(\br) + \hat\Sigma_\nu(\br)
\right ] \xi_\nu(\br) \; d\br \; + \; 
\frac{1}{2} \int \hat\Phi_\nu(\br,\br') \xi_\nu(\br)\xi_\nu(\br')\; 
d\br d\br' \; ,
\ee
in which the integration is over the whole system volume (11) and
$\Sigma_\nu({\bf r})$ is an operator of external fields. If there 
are no external fields, the term $\Sigma_\nu({\bf r})$ plays the role of a 
source mimicking the local influence of the separating surface. 
Characterizing just local perturbations near the separating surface, this 
term has negligible influence, as compared to the single-particle bulk term
$\hat{K}_\nu({\bf r})$ and it can be set to zero after calculating observable 
quantities, as it is done in the Bogolubov method of quasi-averages$^{51}$. 
However, it can be essential if there are external fields or the system 
volume is small.

The total system Hamiltonian 
\be
\label{17}
H(\xi) = \bigoplus_\nu H_\nu(\xi)
\ee
is the direct sum of terms (16). Each of the terms $H_\nu(\xi)$ acts on a 
weighted Hilbert space $\mathcal{H}_\nu$ and the total Hamiltonian (17) is 
defined on the fiber space
\be
\label{18}
\cH = \bigotimes_\nu \cH_\nu \;   ,
\ee
which is the tensor product of the weighted Hilbert spaces$^{43}$. According
to the definition of the equimolecular separation$^{46,47}$, preserving the
additivity of extensive quantities, the operators from the algebra of local 
observables are represented as the sums
\be
\label{19}
 \hat A(\xi) = \bigoplus_\nu \hat A_\nu(\xi) \; ,
\ee
with $\hat{A}_\nu(\xi)$ acting on $\mathcal{H}_\nu$ and $\hat{A}(\xi)$, on 
space (18).

In the present section, for generality, we describe the approach that is
valid for quantum systems. As has been mentioned in the Introduction, 
many biological systems, even of rather large sizes, exhibit quantum 
properties$^{36-40}$. Then, in Secs. 3 and 4, we consider a quantum model
illustrating the approach. In Secs. 5 and 6, we explain how the approach
can be used for characterizing the size transformations induced by 
multiscale density fluctuations in classical systems.

\subsection{Averaging over Multiscale Fluctuations}

Any experiment, accomplished over the time period essentially larger than the 
correlation time (4), will produce the results corresponding to the smoothed
picture averaged over many mesoscopic fluctuations that occurred during the 
observation time. This implies that we do not need to follow the complicated
dynamics of all those fluctuations that have happened during the 
quasi-equilibrium stage (9), but we have to understand how to describe the
related averaged picture. Multiscale fluctuations are characterized by a wide
diapason of time and space scales$^{1-7}$. Therefore averaging over them 
requires to consider various possible snap-shot configurations described in 
the previous subsection. In mathematical terms, this means that it is necessary 
to define functional integration over the manifold indicator functions (14).
Such a functional integration has been rigorously defined in the previous
papers$^{41-43,52,53}$, where all mathematical details can be found. Not 
going into these details, we denote the differential measure of the functional 
integration over sets (15) of manifold indicator functions (14) as 
$\mathcal{D}\xi$. 

First of all, we have to understand how the functional integration over 
sets (15) can be realized for the polynomial functionals of the form
\be
\label{20}
C_n(\xi) = \sum_{\nu_1} \sum_{\nu_2} \ldots
\sum_{\nu_n} \int C_{\nu_1\nu_2\ldots\nu_n}(\br_1,\br_2,\ldots,\br_n)
\xi_{\nu_1}(\br_1) \xi_{\nu_2}(\br_2) \ldots \xi_{\nu_n}(\br_n) \;
d\br_1 d\br_2 \ldots d\br_n \;   ,
\ee
where the integration is over the system volume (11). Here the kernel
$C_{\nu_1\nu_2\ldots\nu_n}(\cdot)$ is an arbitrary operator function. 
When functional (20) represents observable quantities, the kernel 
function is self-adjoint.  

The following proposition$^{42,43,52,53}$ is valid.
 
\vskip 2mm

{\bf Theorem 1}. The functional integration of the polynomial 
functional (20) over the manifold indicator functions (14) gives
\be
\label{21}
 \int C_n(\xi) \; \cD\xi = C_n(w) \;  ,
\ee
where
\be
\label{22}
 C_n(w) = \sum_{\nu_1} \sum_{\nu_2} \ldots
\sum_{\nu_n} w_{\nu_1} w_{\nu_2} \ldots w_{\nu_n}
\int C_{\nu_1\nu_2\ldots\nu_n}(\br_1,\br_2,\ldots,\br_n) \;
d\br_1 d\br_2 \ldots d\br_n \;  ,
\ee
and 
\be
\label{23}
w_\nu \equiv \frac{1}{V} \int \xi_\nu(\br) \; d\br
\ee
is the geometric weight of the $\nu-$phase. \; $\Box$

\vskip 2mm

The operators of observable quantities act on the Hilbert fiber space (18). 
Each of the fiber sections $\mathcal{H}_\nu$ contains information on the 
degrees of freedom corresponding to microscopic atomic fluctuations. The
variables corresponding to mesoscopic multiscale fluctuations are 
represented by the manifold indicator functions (14). A macromolecular 
system, with these two types of degrees of freedom is characterized by the
Gibbs quasi-equilibrium ensemble$^{46,47}$. It is possible to introduce
temperature $T$ as a thermodynamic parameter corresponding to the temperature 
of thermostat surrounding the given system. In a more general interpretation,
temperature can be associated with the intensity of noise, caused by random 
external perturbations$^{29}$. The related thermodynamic potential reads as
\be
\label{24}
 F = - T \ln \int {\rm Tr}_\cH e^{-\bt H(\xi)} \; \cD\xi \; ,
\ee
where the trace is taken over the microscopic degrees of freedom and 
$\beta \equiv 1/T$. Here and everywhere below, we use the system of units 
where the Planck constant and Boltzmann constant are set to one 
($\hbar = 1,\; k_B =1$). The following statement holds$^{42,43,52,53}$.

\vskip 2mm

{\bf Theorem 2}. The integration over the manifold indicator functions (14) 
in the thermodynamic potential (24) gives
\be
\label{25}
F = - T \ln {\rm Tr}_\cH e^{-\bt\widetilde H} =
\sum_\nu F_\nu \; ;
\ee
here the renormalized effective Hamiltonian is
\be
\label{26}
 \widetilde H = \bigoplus_\nu H_\nu \;  ,
\ee
in which
$$
H_\nu = w_\nu \left ( \hat K_\nu + \hat\Sigma_\nu \right )
+ \frac{w^2_\nu}{2} \; \hat\Phi_\nu \; ,
$$
\be
\label{27}
\hat K_\nu \equiv \int \hat K_\nu(\br) \; d\br \; , \qquad
\hat\Sigma_\nu \equiv \int \hat\Sigma_\nu(\br) \; d\br \; ,
\qquad
\hat \Phi_\nu \equiv \int \hat\Phi_\nu(\br,\br') \; d\br d\br' \; ,
\ee
the partial thermodynamic potentials are
\be
\label{28}
 F_\nu = - T \ln {\rm Tr}_{\cH_\nu} e^{-\bt H_\nu}  \; ,
\ee
and the geometric weights $w_\nu$ are the minimizers of the 
thermodynamic potential (25),
\be
\label{29}
F = {\rm abs} \min_{\{w_\nu\} } F(\{ w_\nu\} ) \; ,
\ee
under the conditions
\be
\label{30}
 \sum_\nu w_\nu = 1 \; , \qquad 0 \leq w_\nu \leq 1 \; ,
\ee
defining the set $\{w_\nu\}$ of the geometric weights as a probability 
measure. \; $\Box$

\vskip 2mm

The observable quantities are given by the expectation values of 
operators (19) from the algebra of local observables. These expectation values 
\be
\label{31}
\lgl \hat A \rgl = \int {\rm Tr}_\cH \hat\rho(\xi) \hat A(\xi) \; \cD \xi
\ee
contain the trace over the microscopic degrees of freedom and the averaging
over multiscale mesoscopic fluctuations with the statistical operator
$$
\hat\rho(\xi) \equiv 
\frac{\exp\{-\bt H(\xi\} }{\int{\rm Tr}_\cH\exp\{-\bt H(\xi)\} \cD \xi} \; . 
$$
It is possible to prove the following
theorem$^{42,43,52,53}$.  

\vskip 2mm

{\bf Theorem 3}. Integration over the manifold indicator functions in 
the expectation value (31), for large $N \gg 1$, yields  
\be
\label{32}
 \lgl \hat A \rgl = \sum_\nu \lgl \hat A_\nu \rgl  \; ,
\ee
with the terms
\be
\label{33}
\lgl \hat A_\nu \rgl \equiv  
{\rm Tr}_{\cH_\nu} \hat\rho_\nu \hat A_\nu \;   ,
\ee
partial statistical operators
\be
\label{34}
 \hat\rho_\nu \equiv 
\frac{\exp(-\bt H_\nu)}{{\rm Tr}_{\cH_\nu}\exp(-\bt H_\nu)} \; ,
\ee
and the notation
\be
\label{35}
 \hat A_\nu \equiv \lim_{\{\xi_\nu\ra w_\nu\}} \hat A_\nu(\xi) 
\ee
implying that all $\xi_\nu$ are replaced by $w_\nu$. \; $\Box$

\vskip 2mm

Sometimes, it is more convenient to resort to the equivalent notation
\be
\label{36}
 \lgl \hat A_\nu \rgl \equiv \lgl \hat A \rgl_\nu \;  ,
\ee
which we shall use when appropriate. 

Among the observable quantities, there are the order operators 
$\hat{\eta}_\nu$, whose averages define the order parameters
\be
\label{37}
\eta_\nu = \lgl \hat\eta_\nu \rgl
\ee
allowing us to distinguish different phases.

As a particular example, let us consider the case of two phases, when 
$\nu = 1,2$. Then it is convenient to invoke the notation
\be
\label{38}
 w_1 \equiv w \; , \qquad w_2 = 1 - w \;  ,
\ee
explicitly taking account of normalization (30). The minimization condition
for the thermodynamic potential (25) becomes
\be
\label{39}
 \frac{\prt F}{\prt w} = 0 \; , \qquad
\frac{\prt^2F}{\prt w^2} > 0 \;  .
\ee
The first of the above equations results in
\be
\label{40}
 \left \lgl \frac{\prt\widetilde H}{\prt w} 
\right \rgl = 0  .
\ee
While the inequality in Eq. (39) becomes 
\be
\label{41}
 \left \lgl \frac{\prt^2\widetilde H}{\prt w^2} 
\right \rgl \; > \bt \left\lgl \left (
\frac{\prt\widetilde H}{\prt w} \right )^2 \right \rgl \; .
\ee
In view of the form of Hamiltonian (26), Eq. (40) reduces to
\be
\label{42}
 w = 
\frac{\Phi_2+\widetilde K_2-\widetilde K_1}{\Phi_1+\Phi_2} \;  ,
\ee
where
$$
\widetilde K_\nu = K_\nu + \Sigma_\nu \; , \qquad
K_\nu = \lgl \hat K_\nu \rgl \; ,
$$
\be
\label{43}
 \Sigma_\nu = \lgl \hat\Sigma_\nu \rgl \; , \qquad
\Phi_\nu = \lgl \hat\Phi_\nu \rgl \;  .
\ee
Equation (41) has the meaning of the stability condition and reads as
\be
\label{44}
 \Phi_1 + \Phi_2 > \bt \left\lgl \left (
\frac{\prt\widetilde H}{\prt w} \right )^2 \right \rgl \;  .
\ee
Since the right-hand side here is positive, the necessary stability 
condition is
\be
\label{45}
 \Phi_1 + \Phi_2 > 0 \;  .
\ee

This tells us that multiscale mesoscopic fluctuations arise not in an
arbitrary system, but only in those systems where the stability 
conditions are satisfied. The stability conditions impose restrictions
on the system parameters as well as on the external parameters, defining
the regions of admissible values of these parameters where the multiscale 
fluctuations can exist. The concrete constraints, following from 
inequalities (44) and (45) essentially depend on the considered model. 
The possibility of occurrence of mesoscopic fluctuations has been 
analyzed, for several models of condensed matter, for instance, for 
anharmonic crystals$^{52}$, lattice-gas model$^{54}$, Hubbard model$^{55}$, 
Vonsovsky-Ziener model$^{56}$, and high-temperature superconductors$^{57-59}$. 
Somewhat similar situation happens for the model of mixed nuclear 
matter$^{60,61}$.

\section{Model Macromolecular System}

In the present section, we shall derive a model of a macromolecular system
of $N$ atoms, with $N$ assumed to be large. This model will be used for 
illustrating how the approach of Sec. 2 works. For generality, we consider
a quantum system, since many biological systems to some extent display 
quantum properties$^{36-40}$. Typically, quantum effects in biomolecular 
systems come from fluctuations of protons and long-range electron 
transport/photo excitation in macromolecules. Of course, quantum effects not 
always are important. And in Sections 5 and 6, we show how the approach could 
be used for describing multiscale fluctuations in classical systems. 

Meanwhile, for generality, we consider a generic quantum system described 
by field operators $\psi({\bf r})$. The standard expression for the 
single-particle energy operator is
\be
\label{46}
 \hat K = \int \psi^\dgr(\br) \hat H(\br) \psi(\br) \; d\br \;  ,
\ee
with the single-atom Hamiltonian 
\be
\label{47}
 \hat H(\br) = - \; \frac{\nabla^2}{2m} + U(\br) \;  ,
\ee
where $U({\bf r})$ is an effective potential making the system confined. 
Additional energy, caused by external fields, corresponds to the
operator
\be
\label{48}
\hat\Sigma = \int \psi^\dgr(\br) h(\br) \psi(\br)\; d\br \; .
\ee
And the interaction term has the standard form
\be
\label{49}
 \hat\Phi = \int \psi^\dgr(\br)  \psi^\dgr(\br')\Phi(\br-\br')
\psi(\br')\psi(\br)\; d\br d\br' \; ,
\ee
with $\Phi({\bf r})$ being the pair-interaction potential. 

For what follows, we need a basis of orthonormalized wave functions, 
which can be chosen as the set of the eigenfunctions of the 
Schr\"{o}dinger equation
\be
\label{50}
 \hat H(\br) \psi_n(\br-\br_j) = E_{nj}\psi_n(\br-\br_j) \;  ,
\ee
defining the localized orbitals$^{44}$ centered at ${\bf r_j}$, with 
$j = 1,2, \ldots,N$. Being orthonormalized, these functions enjoy the 
property
$$
 \int \psi_m^*(\br-\br_i) \psi_n(\br-\br_j)\; d\br = \delta_{ij} \delta_{mn} \; .
$$
Because the considered system is confined, its spectrum is discrete.

The field operators can be expanded over the basis of the localized 
orbitals:
\be
\label{51}
\psi(\br) = \sum_{nj} c_{nj} \psi_n(\br-\br_j) \; .
\ee
Substituting this expansion into Eqs. (46) yields the single-atom term
\be
\label{52}
 \hat K = \sum_{nj} E_{nj} c^\dgr_{nj} c_{nj} \; .
\ee
And using the expansion in Eq. (48) results in
\be
\label{53}
 \hat\Sigma = 
\sum_{mn} \sum_{ij} h_{ij}^{mn} c^\dgr_{mi} c_{nj} \;  ,
\ee
where
\be
\label{54}
 h_{ij}^{mn} = \int \psi_m^*(\br-\br_i) h(\br)
\psi_n(\br-\br_j) \; d\br \;  .
\ee
Respectively, for the interaction term (49), we get
\be
\label{55}
 \hat\Phi = \sum_{n_1n_2n_3n_4} \sum_{j_1j_2j_3j_4}
\Phi_{j_1j_2j_3j_4}^{n_1n_2n_3n_4} 
c_{n_1 j_1}^\dgr c_{n_2j_2}^\dgr c_{n_3j_3}c_{n_4j_4} \;  .
\ee

In order to take into account that at each spatial location ${\bf r_j}$
there is just one atom, we impose the unipolarity constraint
\be
\label{56}
 \sum_n c_{nj}^\dgr c_{nj} = 1 \; , \qquad 
c_{mj} c_{nj} = 0 \;  .
\ee
And the condition that atoms are localized reads as
\be
\label{57}
c_{mi}^\dgr c_{nj} = \dlt_{ij} c_{mj}^\dgr c_{nj} \; .
\ee

Conditions (56) and (57) make it straightforward to simplify the above
expressions, so that Eq. (53) reduces to
\be
\label{58}
 \hat\Sigma = \sum_{mn} \sum_j h_{jj}^{mn} c_{mj}^\dgr c_{nj}  
\ee
and Eq. (55) becomes
\be
\label{59}
 \hat\Phi = \sum_{n_1n_2n_3n_4} \sum_{i\neq j} 
V_{ij}^{mnm'n'} c_{mi}^\dgr c_{nj}^\dgr c_{m'j} c_{n'i} \;  ,
\ee
where the notation 
$$
V_{ij}^{mnm'n'} \equiv \Phi_{ijji}^{mnm'n'} \pm \Phi_{ijij}^{mnm'n'}
$$
is used, with the sign plus or minus, depending on Bose or Fermi statistics 
of atoms, respectively.  

Assume that only two lowest energy levels, defined by Eq. (50), are occupied,
so that the index enumerating the levels is $n = 1,2$. The wave functions, 
corresponding to different energy levels, usually possess different symmetry,
which makes zero some of the matrix elements $V_{ij}^{m n m^\prime n^\prime}$.
The remaining matrix elements can be combined into the expressions
$$
A_{ij} \equiv \frac{1}{4} \left ( V_{ij}^{1111} + V_{ij}^{2222} +
2V_{ij}^{1221} \right ) \; , \qquad
B_{ij} \equiv \frac{1}{2} \left ( V_{ij}^{1111} + V_{ij}^{2222} -
2V_{ij}^{1221} \right ) \; ,
$$
\be
\label{60}
 C_{ij} \equiv \frac{1}{2} \left (  V_{ij}^{2222} -V_{ij}^{1111} 
 \right ) \; , \qquad 
I_{ij} \equiv - 2V_{ij}^{1122} \; .
\ee
Also, let us introduce the notations
\be
\label{61}
E_0 \equiv \frac{1}{2N} \sum_j ( E_{1j} + E_{2j} ) \;   ,
\ee
\be
\label{62}
\Om_j \equiv E_{2j} + h_{jj}^{22} - E_{1j} - h_{jj}^{11}
+ \sum_{i(\neq j)} C_{ij} \;   ,
\ee
and
\be
\label{63}
 B_j \equiv - h_{jj}^{12} - h_{jj}^{21} \;  .
\ee

It is convenient to accomplish an operator transformation by introducing
the pseudospin operators 
\be
\label{64}
 S_j^x = \frac{1}{2} \left ( c_{1j}^\dgr c_{1j} - 
c_{2j}^\dgr c_{2j} \right ) \; , \qquad  
 S_j^y = \frac{i}{2} \left ( c_{1j}^\dgr c_{2j} - 
c_{2j}^\dgr c_{1j} \right ) \; , \qquad
 S_j^z = \frac{1}{2} \left ( c_{1j}^\dgr c_{2j} + 
c_{2j}^\dgr c_{1j} \right ) \; .
\ee
These operators enjoy the spin algebra independently from whether the 
considered atoms are bosons or fermions. The inverse transformations are
$$
c_{1j}^\dgr c_{1j} = \frac{1}{2} + S_j^x \; , \qquad
c_{2j}^\dgr c_{2j} = \frac{1}{2} - S_j^x \; ,
$$
\be
\label{65}
 c_{1j}^\dgr c_{2j} = S_j^x - iS_j^y \; , \qquad
c_{2j}^\dgr c_{1j} = S_j^x + i S_j^y \;  .
\ee

By the above definition, the operator $S_j^x$ describes the state imbalance
between the orbitals, the operator $S_j^y$ corresponds to the interstate
current, and $S_j^z$ characterizes transitions between the orbitals.
With these notations, we have
$$
 \hat K + \hat\Sigma + \hat\Phi = N E_0 - \sum_j \left (
\Om_j S_j^x + B_j S_j^z \right ) + \sum_{i\neq j} \left (
\frac{1}{2} \; A_{ij} + B_{ij} S_i^x S_j^x - I_{ij} S_i^z S_j^z 
\right ) \;  .
$$

Note that operators (64) are called pseudospin, since, though they enjoy 
the spin algebra, they do not represent real spins, but just serve as a
convenient mathematical tool characterizing atomic transitions between 
orbitals.

\section{Statistics of Multiscale Fluctuations}

To take into account the influence of multiscale fluctuations, we follow 
the approach presented in Sec. 2. The main order parameter distinguishing
different phases that exhibit multiscale fluctuations is the average
strength of interorbital transitions
\be
\label{66}
s_\nu \equiv \frac{2}{N} \sum_j \lgl S_j^z \rgl_\nu \; .
\ee
This quantity characterizes the intensity of interorbital transitions, or
the strength of interorbital coupling.

Mesoscopic multiscale fluctuations are known to be connected with the 
molecular structure, molecular shape and size$^{1-7}$. These features are
incorporated into the value of the order parameter (66). If there are no 
transitions between the orbitals, then the system is in its ground 
state corresponding to the first orbital with the lowest energy. In the 
majority of applications, the ground-state orbital can be well 
approximated by a Gaussian. But when there appears the coupling between 
the orbitals, the higher orbitals starts contributing to the values of all 
observable quantities. The higher orbitals possess the larger mean radius.
Hence, the involvement of these orbitals implies the increase of the 
system size, such as swelling. Therefore a sharp variation of the order 
parameter (66) is related to an abrupt swelling (or squeezing) of the 
macromolecule. Generally, one can consider a variety of shapes and sizes. 
Here, for simplicity, we take into account two possible phases enumerated 
with $\nu = 1,2$, defining their relation according to the inequality
\be
\label{67}
 s_1 > s_2   
\ee
between their order parameters. 

Here we are developing a general approach for describing statistics of
multiscale fluctuations. And we aim at demonstrating that multiscale 
fluctuations can lead to the occurrence of sharp phase transitions. We 
treat a model system that is convenient for such an illustration. It is 
not our aim to study particular substances, such as viruses. Their 
detailed investigation cannot be done in the frame of one paper, but 
should be a topic for separate publications. 
   
In the case of the macromolecular model of Sec. 3, following the method 
described in Sec. 2, we get the effective renormalized Hamiltonian
\be
\label{68}
 \widetilde H = H_1 \bigoplus H_2 \;  ,
\ee
which takes into account the existence of mesoscopic multiscale fluctuations,
and where
\be
\label{69}
 H_\nu = w_\nu NE_0 - w_\nu \sum_j \left ( \Om_j S_j^x + 
B_j S_j^z \right ) + w_\nu^2 \sum_{i\neq j} \left ( 
\frac{1}{2} \; A_{ij} + B_{ij} S_i^x S_j^x - I_{ij} S_i^z S_j^z 
\right ) \;  .
\ee
According to expressions (60), the value of $B_{ij}$ is much smaller than 
that of $I_{ij}$, hence, can be neglected, to a first approximation. Also,
keeping in mind a large molecule of many atoms, it is admissible to consider
the parameters $\Omega_j$ and $B_j$ as weakly depending on the atomic number,
replacing them by $\Omega$ and $B_0$, respectively.

Accomplishing calculations with Hamiltonian (68), one meets the standard 
problem of the necessity to decouple atomic correlation functions. The simplest
kind of decoupling is the mean-field approximation that, however, does not take
into account the interatomic correlations, which then are completely lost. A
much more elaborated is the Ter Haar approximation$^{62}$ that preserves in full
pair correlations. The disadvantage of the latter approximation is its high
complexity. The intermediate situation is provided by the Kirkwood$^{63}$ 
approximation that is sufficiently simple, at the same time taking account of
atomic correlations. Employing the Kirkwood approximation in our case, we have
\be
\label{70}
 \lgl S_i^\al S_j^\bt \rgl_\nu =
g_{ij}^\nu \lgl S_i^\al \rgl_\nu 
\lgl S_j^\bt \rgl_\nu \;  ,
\ee
where $g_{ij}^\nu$ is a pair correlation function for the $\nu$-phase. In 
terms of operators, acting on the weighted Hilbert space $\mathcal{H}_\nu$, 
this is equivalent to the transformation
\be
\label{71}
 S_i^\al S_j^\bt = g_{ij}^\nu 
\left ( \lgl S_i^\al \rgl_\nu  S_j^\bt +
S_i^\al \lgl S_j^\bt \rgl_\nu   - 
\lgl S_i^\al \rgl_\nu \lgl S_j^\bt \rgl_\nu \right ) \; .
\ee
As is evident, averaging Eq. (71) yields exactly Eq. (70).
 
Implementing the above approximation, we will need the notations
\be
\label{72}
 A \equiv \frac{1}{N} \sum_{i\neq j} A_{ij} \; , \qquad
J \equiv \frac{1}{N} \sum_{i\neq j} I_{ij} \;  .
\ee
Also, it is convenient to define the {\it correlation parameters}
\be
\label{73}
 g_\nu \equiv \frac{1}{N(N-1)} \sum_{i\neq j} g_{ij}^\nu \;  .
\ee
Then, Hamiltonian (69), in the Kirkwood approximation, takes the form
\be
\label{74}
 H_\nu = N \left ( w_\nu E_0 + \frac{w_\nu^2}{2} \; A + 
\frac{w_\nu^2}{4} \; J g_\nu s_\nu^2 \right ) -
\sum_j \left [ w_\nu \Om S_j^x + 
( w_\nu B_0 + w_\nu^2 J g_\nu s_\nu ) S_j^z \right ] \; .
\ee

In order to simplify the formulas, we introduce the dimensionless
reduced Gibbs potentials
\be
\label{75}
G \equiv \frac{F}{NJ} \; , \qquad G_\nu \equiv \frac{F_\nu}{NJ} 
\ee
and the dimensionless quantities
\be
\label{76}
 u \equiv \frac{A}{J} \; , \qquad \om \equiv \frac{\Om}{J} \; , 
\qquad h \equiv \frac{B_0}{J} \;  .
\ee
Also, in the following expressions, temperature will be measured in 
units of $J$. Then for the system Gibbs potential, we get
\be
\label{77}
G = G_1 + G_2 \; , \qquad
G_\nu = e_\nu - T \ln \left ( 2 \cosh \; \frac{w_\nu\Om_\nu}{2T} 
\right ) \;   ,
\ee
where
\be
\label{78}
 e_\nu = w_\nu \; \frac{E_0}{J} + \frac{w_\nu^2}{4} 
\left ( 2u + g_\nu s_\nu^2\right ) \; , \qquad 
\Om_\nu = \sqrt{\om^2+(h+w_\nu g_\nu s_\nu)^2 } \;  .
\ee

Except the order parameters (66), we can calculate the mean orbital 
imbalance
\be
\label{79}
 x_\nu \equiv \frac{2}{N} \sum_j \lgl S_j^x \rgl_\nu \;   .
\ee
The mean interorbital current is found to be zero,
\be
\label{80}
 \lgl S_j^y \rgl = 0 \;  ,
\ee
as it should be for a quasi-equilibrium system. The orbital imbalance (79)
reads as
\be
\label{81}
 x_\nu = \frac{\om}{\Om_\nu} \; 
\tanh \left ( \frac{w_\nu\Om_\nu}{2T} \right ) \;  .
\ee
And for the main order parameters, the orbital coupling (66), we find
\be
\label{82}
  s_\nu = \frac{h+w_\nu g_\nu s_\nu}{\Om_\nu} \;
\tanh \left ( \frac{w_\nu\Om_\nu}{2T} \right ) \; .
\ee
In view of expression (62), the tunneling parameter $\omega$, defined 
in Eq. (76), is small, $\omega \ll 1$, because of which the orbital 
imbalance (81) is always small, $x_\nu \ll 1$. This confirms that the
main order parameter is the mean orbital coupling (66), for which we have
expression (82).

In the Gibbs potential (77), the sum
$$
\sum_\nu w_\nu E_0 = E_0 = const
$$
reduces to a constant, hence can be omitted. Keeping in mind that 
$\omega \ll 1$, we obtain
\be
\label{83}
  G_\nu = \frac{w_\nu^2}{4} \left ( 2u + g_\nu s_\nu^2 \right )
- T \ln \left [ 2 \cosh \left ( \frac{w_\nu h + w_\nu^2 g_\nu s_\nu}{2T} 
\right ) \right ] \; .
\ee
And the orbital-coupling order parameter (82) becomes
\be
\label{84}
  s_\nu = \tanh \left ( \frac{w_\nu h + w_\nu^2 g_\nu s_\nu}{2T} 
\right ) \; .
\ee
  
As in Eq. (38), we set $w_1 \equiv w$. Then, minimizing the Gibbs 
potential (77) with respect to $w$, we obtain
\be
\label{85}
 w = \frac{2u+h(s_1-s_2) -g_2s_2^2}{4u-g_1s_1^2-g_2s_2^2} \;  ,
\ee
which is the representation of form (42) for the considered model.
Expression (85) is, actually, the additional order parameter defining
the geometric weight of the phase with strong orbital coupling. 

Thus, we have the system of interconnected equations for the order parameters 
$s_1, s_2$, and $w$. The phase with the strong orbital coupling is connected 
with essential atomic correlations, because of which the correlation 
parameter (73) for this phase can be set as $g_1 = 1$. While the phase with
the weak orbital coupling is characterized by the correlations only between 
the $z_0$ nearest neighbors, when the correlation parameter (73) is
$$
 g_2 = \frac{z_0}{N} \simeq 0 \qquad ( N \gg 1 ) \;  .
$$

We solve the system of equations for $s_1, s_2$, and $w$ numerically for 
different parameters $u$ and $h$. The parameter $u = A/J$ describes the 
ratio of repulsive to attractive interactions, because of which it can be 
called the {\it repulsion parameter}. And the parameter $h$ measures the
strength of an external field acting on the system. When, solving the 
equations, we get several solutions, then that of them is to be chosen, 
which provides the absolute minimum for the Gibbs potential (77), that is, 
which corresponds to the most stable system. Also, constraint (67),
distinguishing the phases, is to be valid. The latter always holds for all 
$T$ if $h=0$, when $s_2 =0$. If $h$ is finite, there can arise a small region 
of $T \ll h$ in the vicinity of $T=0$, where $s_2$ is not well defined. 
In that case, we define it by an analytical continuation from the case of 
zero to finite $h$, by employing the linear extrapolation typical of numerical 
analysis$^{64}$. Strictly speaking, mesoscopic fluctuations might be absent.
Therefore, looking for the most stable solution, we have to compare the 
Gibbs potential (77), characterizing a macromolecule with multiscale 
fluctuations, and the Gibbs potential $G(w\equiv 1)$ corresponding to a 
macromolecule without mesoscopic fluctuations, when $w \equiv 1$ and there 
is just one order parameter given by the equation
$$
  s_1 = \tanh \left ( \frac{h + s_1}{2T} \right ) \; .
$$
The most stable solution corresponds to the minimal of all those Gibbs
potentials. The results of numerical calculations are presented in 
Figs. 1 to 5.

Depending on the value of the parameter $u$, it is possible to separate 
several regions with qualitatively different behavior. First of all, when
$u \equiv 0$, mesoscopic fluctuations do not arise, as well as they are 
suppressed for very strong external field $h \ra \infty$. Therefore, we 
shall concentrate our attention on the region of $u > 0$ and not too large
external field $h$. We analyze the behavior of the order parameters $w$,
$s_1$, and $s_2$ as functions of $T$ for varying $h$.

In the region $0 < u < 0.5$, the occurrence of multiscale fluctuations
leads to the appearance of first-order phase transitions. Figure 1 shows
the order parameters $w, s_1$, and $s_2$ as functions of temperature $T$
for different values of the field $h$, at fixed $u = 0.3$. At zero $T$, 
there are no mesoscopic fluctuations, but they arise for $T > 0$.

Larger values of the repulsion parameter $u$ make it possible, for some 
fields $h$, the appearance of mesoscopic fluctuations even at $T = 0$. 
This is illustrated in Figs. 2 and 3. The difference of Fig. 3 from 
Fig. 2 is in the existence of a region of external fields, where, instead 
of a first-order phase transition, there occurs a sharp crossover.

When $u > 1.5$, sharp crossovers are presented for low $T$, while the
first-order transitions appear at higher $T$. This is shown in Fig. 4.

The first-order phase transition lines are presented in Fig. 5, as
a function of $h$, at fixed $u$, and as a function of $u$, at given $h$.
Increasing the external field $h$ shifts the phase transition point to the 
right, while increasing $u$ diminishes the transition temperature.
When $u \ra 0$, then the transition temperature tends to infinity, in
agreement with the fact that at $u = 0$, there are no mesoscopic 
fluctuations. For example, at $h = 0.5$ and $u = 0.001$, the transition 
temperature is $T_0 = 626$. 

The general understanding is as follows. Taking into account the existence
of mesoscopic multiscale fluctuations is crucially important. Their 
existence leads to the occurrence of first-order phase transitions or
sharp crossovers between the strong-orbital coupling and weak-orbital 
coupling states of macromolecules. Increasing the intensity of noise,
or temperature, results in the appearance by a jump of mesoscopic 
fluctuations, when the geometric weight of the strong-coupling phase, 
$w$, abruptly falls down. Respectively, the strong-coupling order 
parameter $s_1$ also falls down by a jump, while the weak-coupling order
parameter $s_2$ suddenly increases. The point of such a phase transition   
depends on the values of the system parameters $u$ and $h$.

\section{Classical Multiscale Fluctuations}

As has been mentioned above, we have considered the case of multiscale
fluctuations in a quantum macromolecular system. This has been done 
for generality, keeping in mind that many biological systems possess
quantum properties. However, multiscale fluctuations can equally arise 
in classical systems. In order to show how they should be described in 
that case, we present below the approach for considering the statistics
of multiscale fluctuations in the language of classical statistical 
mechanics.     

Let us consider a system composed of $N$ atoms or molecules, enumerated 
with the index $i = 1,2,\ldots,N$. The collections of all spatial 
variables and momenta are denoted through the sets
\be
\label{86}
 r^N \equiv \{ \br_1,\br_2,\ldots,\br_N\} \; , \qquad 
 p^N \equiv \{ \bp_1,\bp_2,\ldots,\bp_N\} \;  ,
\ee
respectively. The related differentials are abbreviated as
$$
dr^N \equiv \prod_{i=1}^N d\br_i \; , \qquad
dp^N \equiv \prod_{i=1}^N d\bp_i \;   .
$$

Kinetic energy of $N$ atoms is
\be
\label{87}
  K(p^N) = \sum_{i=1}^N \frac{\bp_i^2}{2m} \;  .
\ee
And the potential energy writes as
\be
\label{88}
 \frac{1}{2} \; \Phi\left ( r^N \right ) =
\frac{1}{2} \sum_{i\neq j}^N \Phi(\br_i -\br_j) \;  .
\ee

The overall consideration is the same as in Sec. 2. Under a fixed 
spatial separation of phases, we have the Hamiltonian 
$H(\xi) = \sum_\nu H_\nu$, with the terms 
\be
\label{89}
 H_\nu \left ( \xi,r^N,p^N\right ) = 
\sum_{i=1}^N \xi_{\nu i} \; \frac{\bp_i^2}{2m} +
\frac{1}{2} \sum_{i\neq j}^N \xi_{\nu i} \xi_{\nu j} 
\Phi(\br_i -\br_j) \; ,
\ee
where $\xi_{\nu i} \equiv \xi_\nu({\bf r}_i)$ is the manifold indicator
defined in Eq. (14). The latter is normalized as
\be
\label{90}
 \sum_{i=1}^N \xi_\nu(\br_i) = N_\nu \; ,
\ee
with $N_\nu$ being the number of atoms composing the $\nu$-th phase. 
Accomplishing the averaging over the configurations of multiscale
fluctuations, according to Sec. 2, we get the effective Hamiltonian,
similar to Eq. (26), with the terms
\be
\label{91}
 H_\nu\left ( r^N, p^N \right ) =w_\nu K(p^N) +
\frac{w_\nu^2}{2} \; \Phi(r^N ) \;   ,
\ee
where the phase weights are
\be
\label{92}
 w_\nu = \frac{N_\nu}{N} \;  .
\ee

The canonical partition function for a heterophase system with 
multiscale fluctuations reads as
\be
\label{93}
 Z(T,V,N) = \prod_\nu Z_\nu(T,V,N) \; , \qquad
Z_\nu(T,V,N) = \int_{\cal{G}} \exp\left \{
-\bt H_\nu ( r^N , p^N ) \right \} \; 
\frac{dr^N dp^N}{N!(2\pi)^{3N} } \; ,
\ee
where the integration is over the phase volume
\be
\label{94}
 {\cal{G}} \equiv \mathbb{V}^N \times \mathbb{R}^{3N} =
\{ \br_i\in \mathbb{V} , \; \bp_i \in \mathbb{R}^3: \;
 i = 1,2, \ldots,N \} \;  .
\ee
Integrating out the momenta, we have
\be
\label{95}
 Z_\nu(T,V,N) = \left ( \frac{mT}{2\pi w_\nu} \right )^{3N/2}
\int_{\mathbb{V}^N} \exp \left \{ -\; \frac{w_\nu^2}{2T} \;
\Phi(r^N) \right \} \; \frac{dr^N}{N!} \;  .
\ee

Having defined the system partition function, it is straightforward
to calculate the desired thermodynamic characteristics.

\section{Density and Size Fluctuations}

To be more specific, let us consider the case when multiscale 
fluctuations correspond to mesoscopic density fluctuations, so that,
at each moment of time, the system consists of randomly distributed 
mesoscopic regions having density $\rho_1$, which are intermixed with
the regions of density $\rho_2$. And let, for concreteness, the first 
phase be more dense than the second one:
\be
\label{96}
 \rho_1 > \rho_2 \;  .
\ee
The densities $\rho_\nu$ play here the role of the order parameters 
distinguishing different phases, the dense phase ($\nu = 1$) and the 
rarified phase ($\nu = 2$).

It is clear that, when the majority of $N$ atoms are in the dense 
phase, the system volume $V_1 \sim N/\rho_1$ is smaller than the 
volume $V_2 \sim N/\rho_2$, when almost all atoms are in the rarified 
phase. The transition from the smaller volume $V_1$ to the larger 
volume $V_2$ implies the system swelling, while the opposite transition 
from the larger volume $V_2$ to the smaller volume $V_1$ describes the 
system squeezing. 

Since in this consideration, the system volume is not conserved, it 
is necessary to employ the isobaric ensemble, for which the system 
partition function is a function of temperature $T$, pressure $P$,
and the total number of atoms $N$. For the heterophase system, the
isobaric partition function is
\be
\label{97}
 Q(T,P,N) = \prod_\nu Q_\nu(T,P,N) \; , \qquad
Q_\nu(T,P,N) = \int_0^\infty e^{-\bt PV} Z_\nu(T,V,N) \; dV \; .
\ee
So that the Gibbs potential is the sum
\be
\label{98}
 G(T,P,N) =  \sum_\nu G_\nu(T,P,N)
\ee
of the terms
\be
\label{99}
  G_\nu(T,P,N) = - T \ln  Q_\nu(T,P,N) \; .
\ee
 
The phase weights are defined by minimizing the Gibbs potential over
$w_\nu$, under the normalization condition $w_1 + w_2 = 1$. Setting
$w \equiv w_1, w_2 = 1 - w$ makes it possible to write the minimization
condition as
\be
\label{100}
\frac{\prt G}{\prt w} = 0 \; , \qquad G = G(T,P,N) \; .
\ee

The observable quantities can be represented as the averages over the 
distribution function
$$
 f_\nu(r^N,p^N,V) = 
\frac{1}{Q_\nu} \; \exp \{ -\bt (H_\nu + PV ) \} \;  ,
$$
where the Hamiltonian $H_\nu = H_\nu(r^N,p^N)$ is given in Eq. (91). 
This distribution is normalized by the condition
$$
\int_0^\infty \left \{ \int_{\cal{G}} f_\nu(r^N,p^N,V) \;
\frac{dr^N dp^N}{N!(2\pi)^{3N}} \right \} \; dV = 1 \; .
$$
The average of a function $A_\nu = A_\nu(r^N,p^N)$ is defined as
$$
 \lgl A_\nu \rgl = \int_0^\infty  
\left \{ \int_{\cal{G}} A_\nu(r^N,p^N) f_\nu(r^N,p^N,V)\;
\frac{dr^N dp^N}{N!(2\pi)^{3N} } \right \} \; dV \; .
$$

For example, the mean kinetic energy of the $\nu$-th phase, takes 
the form
$$
 K_\nu  = \int_0^\infty  
\left \{ \int_{\cal{G}} K_\nu(p^N) f_\nu(r^N,p^N,V)\;
\frac{dr^N dp^N}{N!(2\pi)^{3N}} \right \} \; dV \;  ,
$$
which can be reduced to 
$$
 K_\nu  = \frac{3TN}{2w_\nu} \;  .
$$
And the mean potential energy of the $\nu$-th phase is $(1/2) \Phi_\nu$,
where 
$$
  \Phi_\nu  = \int_0^\infty  
\left \{ \int_{\cal{G}} \Phi_\nu(p^N) f_\nu(r^N,p^N,V)\;
\frac{dr^N dp^N}{N!(2\pi)^{3N} } \right \} \; dV \;  .
$$

Minimizing the Gibbs potential (98), according to condition (100), 
with taking into account that 
$$
 \frac{\prt G}{\prt w_\nu} = \left\lgl 
\frac{\prt H_\nu}{\prt w_\nu} \right \rgl \;  ,
$$
we find the equation for the weight of the dense phase
$$
 w = \frac{\Phi_2+K_2-K_1}{\Phi_1+\Phi_2} \;  ,
$$
which is similar to Eq. (42).  

The total system volume is defined through the derivative
$$
 V(T,P,N) = \frac{\prt G}{\prt P} = V_1 + V_2 \;  ,
$$
with the partial volumes 
$$
 V_\nu = \frac{\prt G_\nu}{\prt P} = V_\nu(T,P,N) \;  .
$$
In this way, we find the densities of the competing phases
$$
\rho_\nu \equiv \frac{N_\nu}{V_\nu} =
w_\nu N \left ( \frac{\prt G_\nu}{\prt P} \right )^{-1} \;  .
$$

At low temperature, when $w_1 > w_2$, almost all atoms are in the 
dense phase, so that $N \sim N_1$ and $V \sim V_1$, though the 
occurrence of the multiscale density fluctuations results in the 
existence of the regions of the rarified phase. Rising temperature
leads to the intensification of the mesoscopic density fluctuations
and to the increase of the weight $w_2$. Then, these multiscale 
density fluctuations provoke the transition to the state, where 
$w_2 > w_1$, so that the system volume increases, becoming close 
to $V_2$, though a small admixture of the dense-phase fluctuations 
can remain. Such a swelling transition is illustrated by Fig. 6.

\section{Conclusion}

Macromolecular systems are investigated experiencing, in addition to fast 
atomic fluctuations, slow multiscale fluctuations corresponding to the 
coherent motion of groups of many atoms. In the process of this motion, 
the system symmetry can be locally broken or restored in a spontaneous 
way$^{65}$. When the experimental observation time of the system is 
sufficiently long, being much longer than the atomic correlation time, 
one does not need to study the detailed dynamics of all atoms. In such 
experiments, one observes the smoothed picture averaged over many 
mesoscopic multiscale fluctuations. Then what one has to know is the 
statistics of multiscale fluctuations, that is, the average influence 
of these fluctuations on the values of observable quantities.  

A method is suggested for treating the statistical influence of multiscale 
fluctuations on observables. The method is based on the Gibbs theory of 
quasi-equilibrium systems, with averaging over the ensemble of multiscale 
fluctuations. As a result, one comes to an effective statistical system
with renormalized interactions and with the influence of the multiscale 
fluctuations incorporated into geometric weights of competing phases.

The method is illustrated for a macromolecular system exhibiting two 
possible states differing by the strength of coupling between atomic 
orbitals. Multiscale fluctuations in space and time occur between these 
two states, of weak-orbital and strong-orbital coupling. The order parameters 
of the system are calculated. The presence of mesoscopic fluctuations is 
shown to lead to sharp phase transitions between the phases of orbital 
weak-coupling and strong-coupling. Such phase transitions can be 
accompanied by an essential swelling or squeezing of macromolecules. 

As is shown, the existence of mesoscopic multiscale fluctuations may result 
in drastic changes in the properties of macromolecules. When varying 
temperature, or other thermodynamic parameters, macromolecules can change 
their properties not gradually, but in a sharp phase transition. This is 
necessary to take into account in analyzing the behavior of macromolecules
under the variation of external conditions. For example, considering the 
problem of the origin of life on Earth, one assumes that life appeared
after the primordial ocean was cooled down $^{66-70}$. In this respect, 
one often argues that it looks strange that biomolecules have arisen so 
suddenly, while the process of cooling was gradual. However, this does not 
seem strange if we take into account mesoscopic multiscale fluctuations. 
As is shown above, under a slow gradual cooling, the properties of 
macromolecules can change by an abrupt jump. It is, probably, the 
occurrence of mesocopic fluctuations that facilitated the spontaneous 
generation of life on Earth. 

The aim of the present paper has been threefold. First, to develop a 
general approach allowing one to describe the statistics of multiscale 
fluctuations in macromolecules, without the need of studying their 
dynamics. Second, to demonstrate the use of the approach by a model that, 
though being simple, nevertheless exhibits rather rich properties, with 
phase transitions provoked by the multiscale fluctuations. Third, to 
emphasize that the approach is equally applicable to quantum as well 
as to classical systems. We have limited ourselves by these principal 
problems. Applications to particular biomolecules, having complicated 
structure, require separate investigations, with extensive computer 
calculations that are out of the scope of the present paper and are 
planned for future research. 

\vskip 3mm

{\bf Acknowledgement}

The authors acknowledge financial support from the Russian Foundation
of Basic Research.

\newpage

{\bf References}
\vskip 5mm

{\parindent = 0pt
(1) Miao, Y.; Ortoleva, P.J. 
{\it J. Chem. Phys.} {\bf 2006}, {\it 125}, 214901. 

(2) Pankavich, S.,; Mao, Y.; Ortoleva, J.; Shreif, Z.; Ortoleva, P.
{\it J. Chem. Phys.} {\bf 2008}, {\it 128}, 234908.

(3) Pankavich, S.; Shreif, Z.; Ortoleva, P. 
{\it Physica A} {\bf 2008}, {\it 387}, 4053-4069.

(4) Mao, Y.; Johnson, J.E.; Ortoleva, P.J. 
{\it J. Phys. Chem. B} {\bf 2010}, {\it 114}, 11181-11195.

(5) Pankavich, S.; Ortoleva, P. 
{\it J. Math. Phys.} {\bf 2010}, {\it 51}, 063303.

(6) Singharoy, A.; Cheluvaraja, S.; Ortoleva, P.
{\it J. Chem. Phys.} {\bf 2011}, {\it 134}, 044104.

(7) Shreif, Z.; Ortoleva, P. 
{\it J. Chem. Phys.} {\bf 2011}, {\it 134}, 104106.

(8) Wernsdorfer, W. 
{\it Adv. Chem. Phys.} {\bf 2001}, {\it 118}, 99-190.

(9) Ferr\'{e}, J. 
{\it Topics Appl. Phys.} {\bf 2002}, {\it 83}, 127-168.

(10) Yukalov, V.I.; Yukalova, E.P. 
{\it Phys. Part. Nucl.} {\bf 2004}, {\it 35}, 348-382.

(11) Liu, J.P.; Fullerton, E.; Gutfleisch, O.; Sellmyer, D.J., Eds. 
{\it Nanoscale Magnetic Materials and Applications}; Springer: Berlin, 2009.

(12) Chakraborty, T. {\it Quantum Dots}; North Holland: Amsterdam, 1999.

(13) Leatherdale, C.A.; Woo, W.K.; Mikulec, F.V.; Bawendi, M.G. 
{\it J. Phys. Chem. B} {\bf 2002}, {\it 106}, 7619–7622.

(14) Lieb, E.H.; Seiringer, R.; Solovej, J.P.; Yngvason, J.
{\it The Mathematics of the Bose Gas and Its Condensation};
Birkhauser: Basel, 2005.
 
(15) Letokhov, V. {\it Laser Control of Atoms and Molecules}; 
Oxford University: New York, 2007.

(16) Pethik C.J.; Smith, H. 
{\it Bose-Einstein Condensation in Dilute Gases}; 
Cambridge University: Cambridge, 2008.

(17) Courteille, P.W.; Bagnato, V.S.; Yukalov, V.I. 
{\it Laser Phys.} {\bf 2001}, {\it 11}, 659-800.

(18) Andersen, J.O. 
{\it Rev. Mod. Phys.} {\bf 2004}, {\it 76}, 599- 639.

(19) Yukalov, V.I. 
{\it Laser Phys. Lett.} {\bf 2004}, {\it 1}, 435-461.

(20) Bongs, K.; Sengstock, K. 
{\it Rep. Prog. Phys.} {\bf 2004}, {\it 67}, 907-963.

(21) Yukalov, V.I.; Girardeau, M.D. 
{\it Laser Phys. Lett.} {\bf 2005}, {\it 2}, 375-382.

(22) Posazhennikova, A. 
{\it Rev. Mod. Phys.} {\bf 2006}, {\it 78}, 1111-1134.

(23) Yukalov, V.I. 
{\it Laser Phys. Lett.} {\bf 2007}, {\it 4}, 632-647.

(24) Proukakis, N.P.; Jackson, B. 
{\it J. Phys. B} {\bf 2008}, {\it 41}, 203002. 

(25) Yurovsky, V.A.; Olshanii, M.; Weiss, D.S. 
{\it Adv. At. Mol. Opt. Phys.} {\bf 2008}, {\it 55}, 61-138.

(26) Ketterle, W.; Zwierlein, M.W. 
{\it Riv. Nuovo Cimento} {\bf 2008}, {\it 31}, 247-422.

(27) Moseley, C.; Fialko, O.; Ziegler, K. 
{\it Ann. Phys. (Berlin)} {\bf 2008}, {\it 17}, 561-608.

(28) Yukalov, V.I. 
{\it Laser Phys}. {\bf 2009}, {\it 19}, 1-110.

(29) Yukalov, V.I. 
{\it Phys. Part. Nucl.} {\bf 2011}, {\it 42}, 460-513.

(30) Lee, S.W.; Mao, C.; Flynn, C.E.; Belcher, A.M. 
{\it Science} {\bf 2002}, {\it 296}, 892-895.

(31) Frenkel, J. {\it Kinetic Theory of Liquids}; Clarendon: Oxford, 1946.

(32) Fisher, M.E. 
{\it Nature of Critical Points}; Colorado University: Boulder, 1965.

(33) Fisher, M.E. 
{\it Rep. Prog. Phys.} {\bf 1967}, {\it 30}, 615-730.

(34) Khait, Y.L. 
{\it Phys. Rep.} {\bf 1983}, {\it 99}, 237-340.

(35) Khait, Y.L. 
{\it Atomic Diffusion in Solids}; Scitec: Zürich, 1997.

(36) Engel, G.S.; Calhoun, T.S.; Read, E.L.; Ahn, T.K.; Mancal, T.; 
Cheng, Y.C.; Blankenship, R.E.; Fleming, G.R.
{\it Nature} {\bf 2007}, {\it 446}, 782-786. 

(37) Collini, E.; Wong, C.Y.; Wilk, K.E.; Curni, P.M.; Bruner, P.; 
Scholes, G.D.
{\it Nature} {\bf 2010}, {\it 463}, 644-647. 

(38) Panitchayangkoon, G.; Hayes, D.; Fransted, K.A.; Caram, J.R.; 
Harel, E.; Wen, J.; Blankenship, R.E.; Engel, G.S.
{\it Proc. Natl. Acad. Sci. USA} {\bf 2010}, {\it 107}, 12766-12770.  

(39) Gauger, E.M.; Rieper, E.; Morton, J.J.; Benjamin, S.C.; Vedral, V.
{\it Phys. Rev. Lett.} {\bf 2011}, {\it 106}, 040503.

(40) Franco, M.I.; Turin, L.; Mershin, A.; Skoulakis, E.M.
{\it Proc. Natl. Acad. Sci. USA} {\bf 2011}, {\it 108}, 3797-3802.

(41) Yukalov, V.I., {\it Physica A} {\bf 1986} {\it 136}, 575-587.

(42) Yukalov, V.I. {\it Physica A} {\bf 1987}, {\it 141}, 352-374.

(43) Yukalov, V.I. {\it Phys. Rep.} {\bf 1991}, {\it 208}, 395-489.
 
(44) Coleman, A.J.; Yukalov, V.I. {\it Reduced Density Matrices};
Springer: Berlin, 2001.

(45) Yukalov, V.I. {\it Physica A} {\bf 2002}, {\it 310}, 413-434. 

(46) Gibbs, J.W. {\it Collected Works}; Longmans: New York, 1928. 

(47) Gibbs, J.W. {\it Elementary Priciples in Statistical Mechanics};
Cambridge University: Cambridge, 2010.

(48) Ono, S.; Kondo, S. {\it Molecular Theory of Surface Tension in Liquids};
Springer: Berlin, 1960.

(49) Krotov, V.V.; Rusanov, A.I. {\it Physicochemical Hydrodynamics of 
Capillary Systems}; World Scientific: Singapore, 1999.

(50) Terletsky, Y.P. {\it Statistical Physics}; Higher School: Moscow, 1973.

(51) Bogolubov, N.N. {\it Lectures on Quantum Statistics}, Vol. 2;
Gordon and Breach: New York, 1970.

(52) Yukalov, V.I. {\it Phys. Rev. B} {\bf 1985}, {\it 32}, 436-446.

(53) Yukalov, V.I. {\it Phys. Lett. A} {\bf 1987}, {\it 125}, 95-100.

(54) Yukalov, V.I.; Yukalova, E.P. {\it Int. J. Mod. Phys. B} {\bf 2001},
{\it 15}, 2433-2453.

(55) Boky, M.A.; Kudryavtsev, I.K.; Yukalov, V.I. {\it Solid State Commun.}
{\bf 1987}, {\it 63}, 731-735.

(56) Boky, M.A.; Yukalov, V.I. in {\it Problems of Statistical Mechanics},
Bogolubov, N.N. Ed., Vol. 1, p. 170-176; JINR: Dubna, 1984.  

(57) Yukalov, V.I. {\it Int. J. Mod. Phys. B} {\bf 1992}, {\it 6}, 91-107.

(58) Coleman, A.J.; Yukalova, E.P.; Yukalov, V.I. 
{\it Physica C}, {\bf 1995}, {\it 243}, 76-92. 

(59) Yukalov, V.I.; Yukalova, E.P. {\it Phys. Rev. B} {\bf 2004}, 
{\it 70}, 224516.

(60) Yukalov, V.I.; Yukalova, E.P. 
{\it Physica A} {\bf 1997}, {\it 243}, 382-414.

(61) Yukalov, V.I.; Yukalova, E.P. 
{\it Phys. Part. Nucl.} {\bf 1997}, {\it 28}, 37-65.

(62) Ter Haar, D. {\it Lectures on Selected Topics in Statistical Mechanics};
Pergamon: Oxford, 1977.

(63) Kirkwood, J.G. {\it Quantum Statistics and Cooperative Phenomena};
Gordon and Breach: New York, 1965.

(64) Mitin, A.V.; Hirsch, G. {\it J. Math. Chem.} {\bf 1994}, {\it 15}, 109-113.

(65) Yukalov, V.I. {\it Phys. Lett. A} {\bf 1981}, {\it 85}, 68-71.

(66) Oparin, A.I. {\it Origin of Life}; Dover: New York, 1953. 

(67) Bernal, J.D. {\it Origins of Life}; Wiedenfeld and Nicholson: London, 1969.

(68) Dyson, F. {\it Origins of Life}; Cambridge University: Cambridge, 1985. 

(69) Bryson, B. {\it A Short History of Nearly Everything}; Black Swan: London, 2004. 

(70) Egel, R.; Lankenau, D.H.; Mulkidjanian, A.Y. {\it Origins of Life: 
The Primal Self-Organization}; Springer: Berlin, 2011. 
}

\newpage

{\Large{\bf Figure Captions} }

\vskip 5mm

Fig. 1. Macromolecule order parameters: (a) geometric weight of the 
strong-coupling phase $w$; (b) strong-coupling phase order parameter 
$s_1$; (c) weak-coupling phase order parameter $s_2$. Dependence on 
dimensionless temperature is shown for the repulsion parameter $u = 0.3$ 
and varying strength of external field: (1) $h = 0.01$; (2) $h = 0.1$; 
(3) $h = 0.2$; (4) $h = 0.3$; (5) $h = 0.5$; (6) $h = 1$. The corresponding 
first-order transition temperatures are: (1) $T_0= 0.145$; (2) $T_0 =0.241$; 
(3) $T_0 = 0.336$; (4) $T_0 = 0.436$; (5) $T_0 = 0.660$; (6) $T_0 = 1.422$.  

\vskip 5mm
Fig. 2. Macromolecule order parameters: (a) $w$; (b) $s_1$; (c) $s_2$. 
Dependence on dimensionless temperature is shown for the repulsion parameter 
$u = 0.6$ and varying strength of external field: (1) $h = 0.01$; (2) $h = 0.1$;
(3) $h = 0.2$; (4) $h = 0.3$; (5) $h = 0.5$; (6) $h = 1$. The corresponding 
transition temperatures are: (1) $T_0= 0.164$; (2) $T_0 =0.217$; 
(3) $T_0 = 0.276$; (4) $T_0 = 0.326$; (5) $T_0 = 0.460$; (6) $T_0 = 0.877$.  

\vskip 5mm
Fig. 3. Macromolecule order parameters: (a) $w$; (b) $s_1$; (c) $s_2$. 
Dependence on dimensionless temperature is shown for $u = 0.8$ and varying 
external field: (1) $h = 0.01$; (2) $h = 0.1$; (3) $h = 0.2$; (4) $h = 0.3$; 
(5) $h = 0.5$; (6) $h = 0.8$; (7) $h = 1$; (8) $h = 2$. The first-order 
transition temperatures are: (1) $T_0 = 0.143$; (7) $T_0 = 0.715$; 
(8) $T_0 = 1.764$. For the values $0.01 < h < 1$, the transition is 
a sharp crossover.

\vskip 5mm
Fig. 4. Macromolecule order parameters: (a) $w$; (b) $s_1$; (c) $s_2$. 
Dependence on dimensionless temperature is shown for $u = 1.5$ and varying 
external field: (1) $h = 0.01$; (2) $h = 0.5$; (3) $h = 1$; (4) $h = 2$; 
(5) $h = 2.5$; (6) $h = 3$; (7) $h = 5$; (8) $h = 6$. The transition 
temperatures are: (7) $T_0 = 4.547$; (8) $T_0 = 6.4$. For the values 
$h < 5$, the transition occurs as a sharp crossover. 

\vskip 5mm
Fig. 5. First-order transition temperature $T_0$: (a) as a function of 
$h$ at fixed $u = 0.3$; (b) as a function of $u$ at fixed $h = 0.5$. 

\vskip 5mm
Fig. 6. Qualitative illustration of the macromolecule swelling provoked 
by multiscale mesoscopic density fluctuations.

\newpage

\begin{figure}[ht]
\vspace{9pt}
\centerline{
\hbox{ \includegraphics[width=8.5cm]{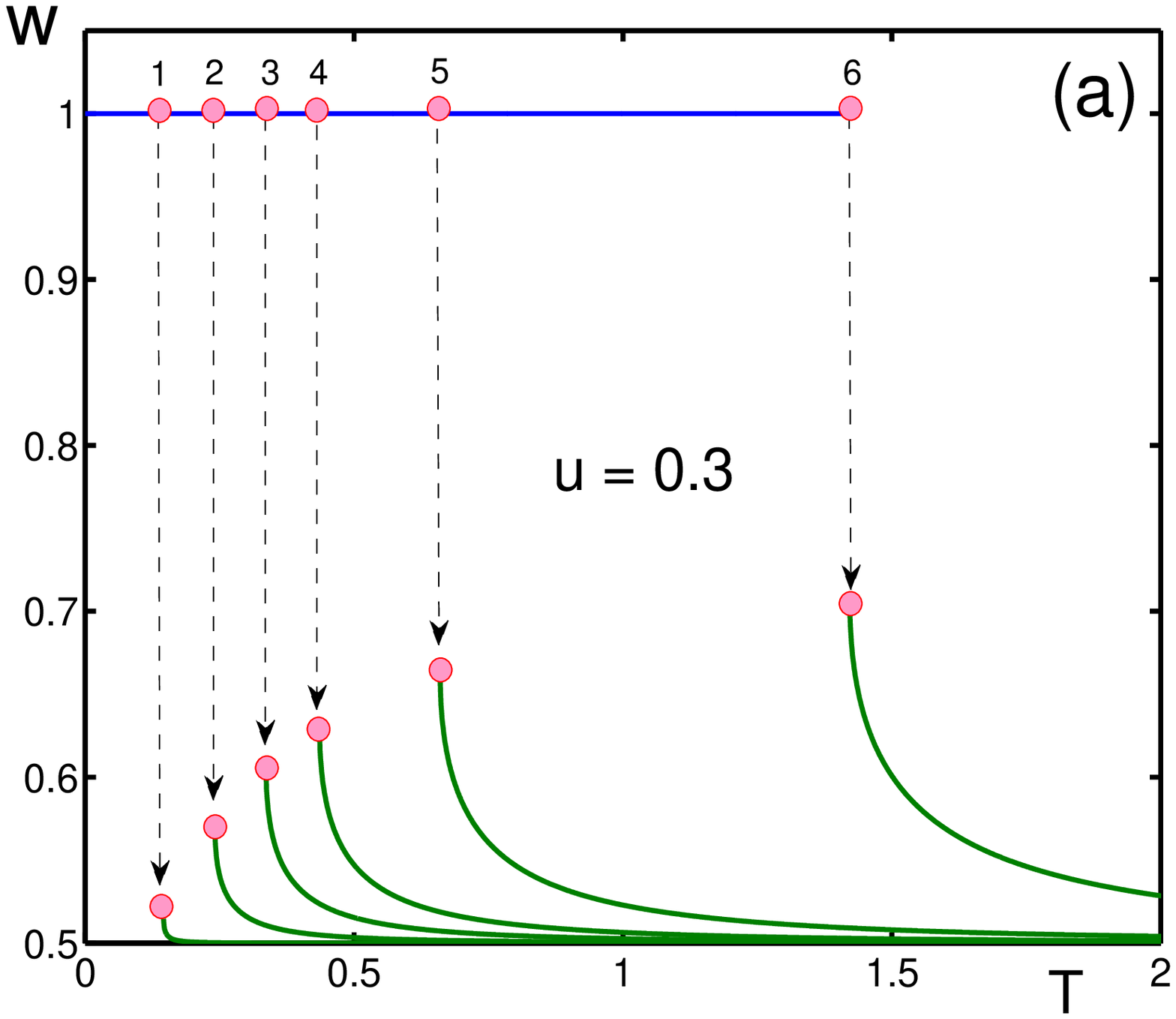} \hspace{2cm}
\includegraphics[width=8.5cm]{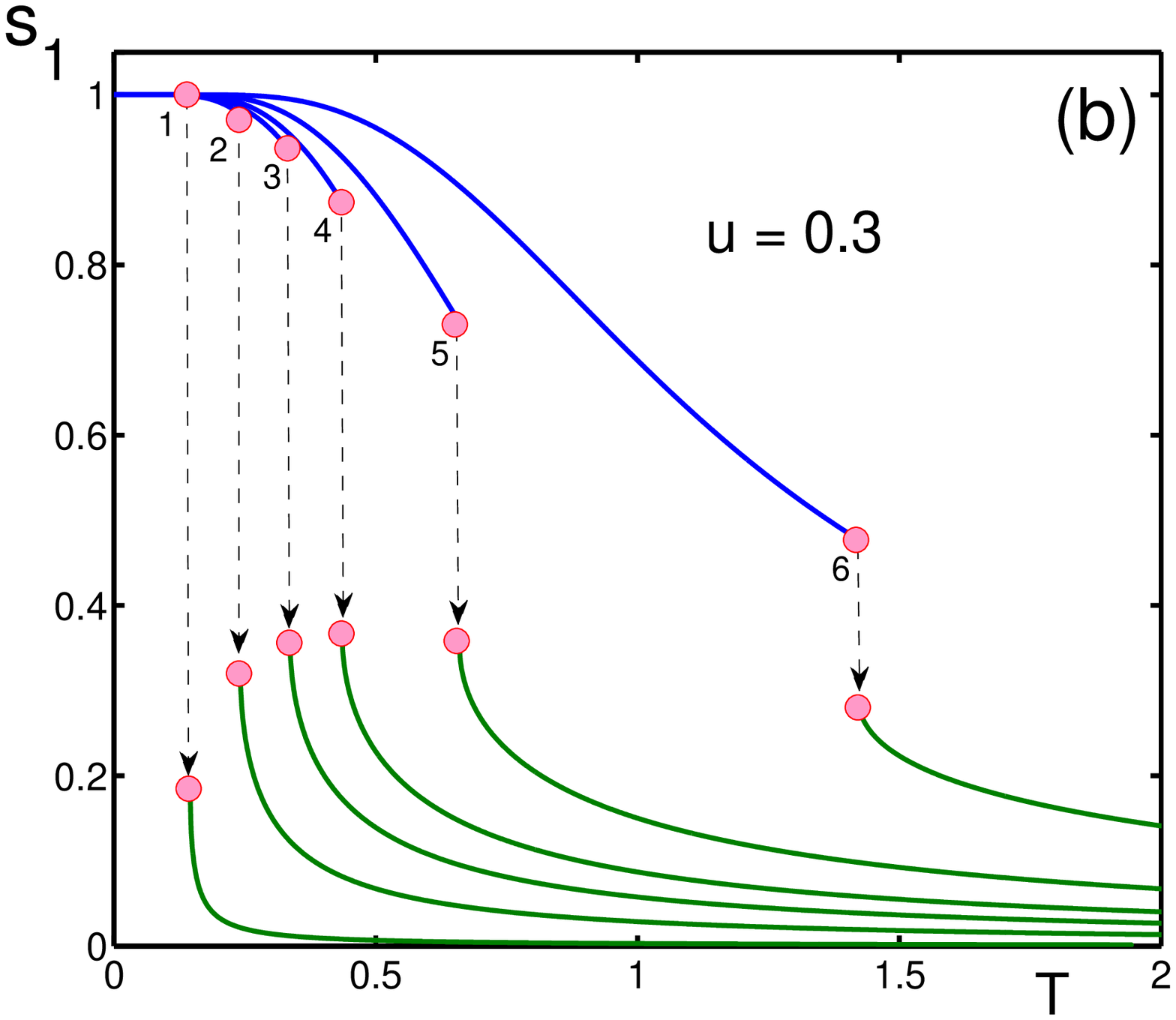} } }
\vspace{9pt}
\centerline{
\hbox{ \includegraphics[width=8.5cm]{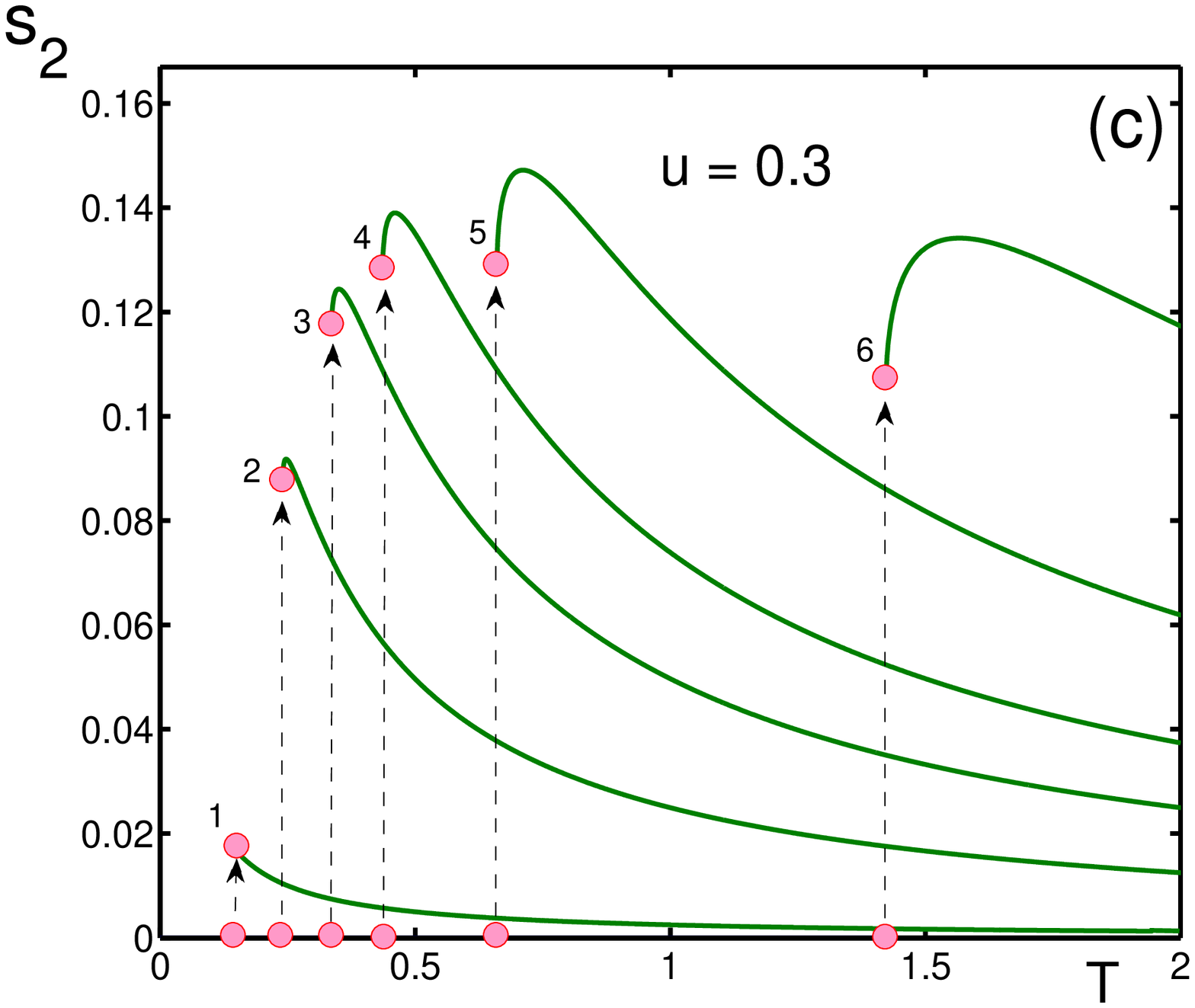}  } }
\caption{ Macromolecule order parameters: (a) geometric weight of the 
strong-coupling phase $w$; (b) strong-coupling phase order parameter 
$s_1$; (c) weak-coupling phase order parameter $s_2$. Dependence on 
dimensionless temperature is shown for the repulsion parameter $u = 0.3$ 
and varying strength of external field: (1) $h = 0.01$; (2) $h = 0.1$; 
(3) $h = 0.2$; (4) $h = 0.3$; (5) $h = 0.5$; (6) $h = 1$. The corresponding 
first-order transition temperatures are: (1) $T_0= 0.145$; (2) $T_0 =0.241$; 
(3) $T_0 = 0.336$; (4) $T_0 = 0.436$; (5) $T_0 = 0.660$; (6) $T_0 = 1.422$. }
\label{fig:Fig.1}
\end{figure}

\begin{figure}[ht]
\vspace{9pt}
\centerline{
\hbox{ \includegraphics[width=8.5cm]{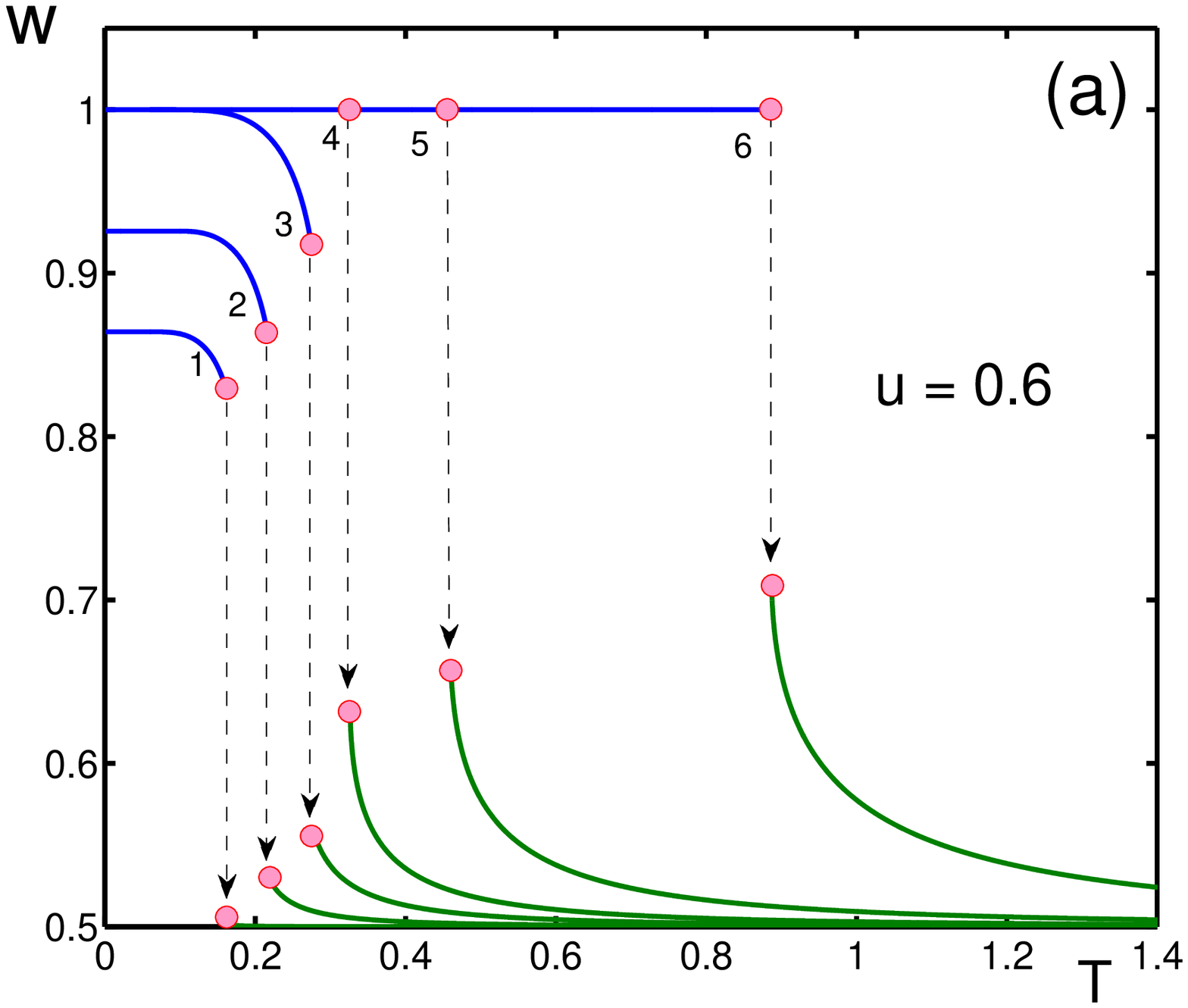} \hspace{2cm}
\includegraphics[width=8.5cm]{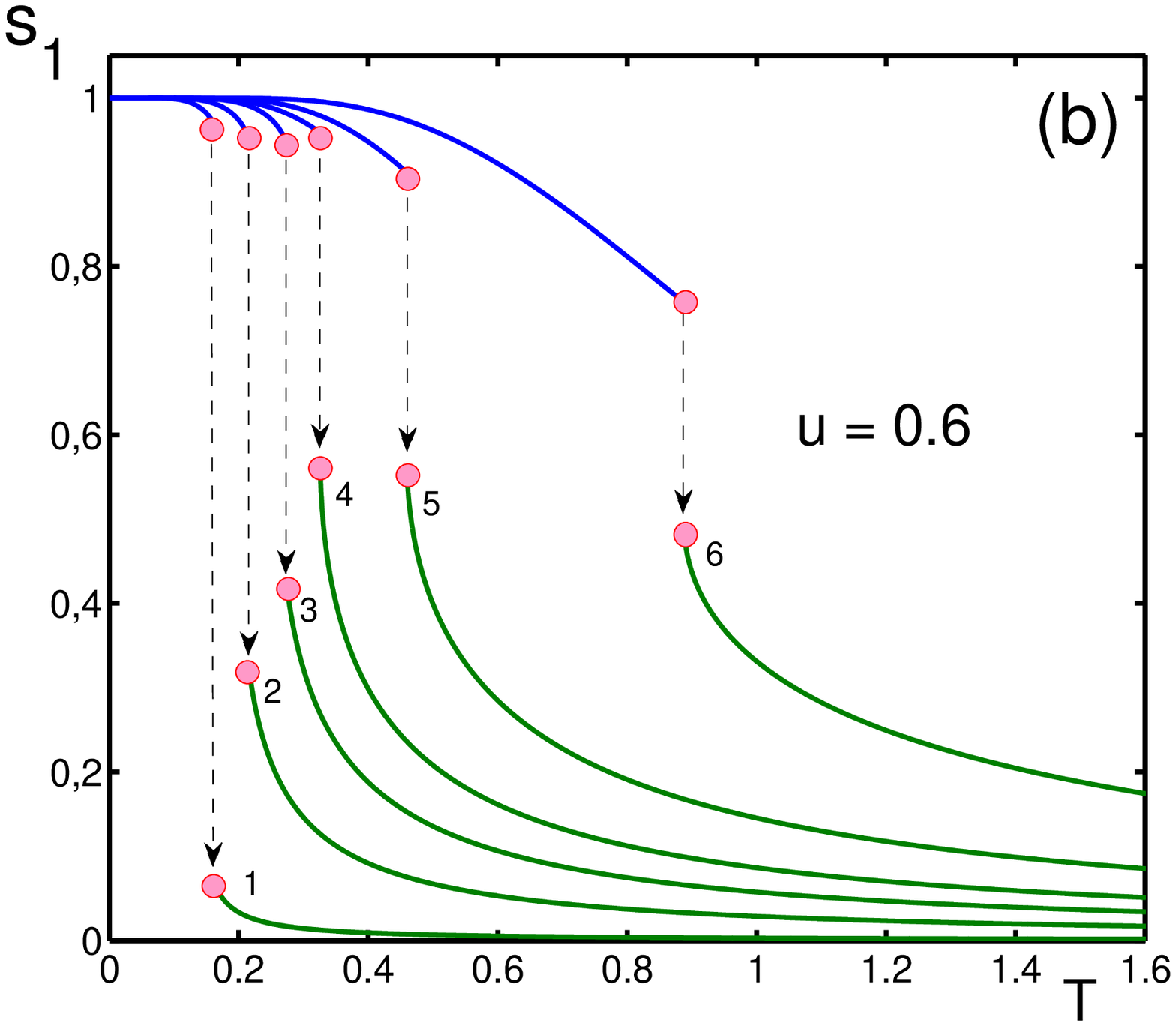} } }
\vspace{9pt}
\centerline{
\hbox{ \includegraphics[width=8.5cm]{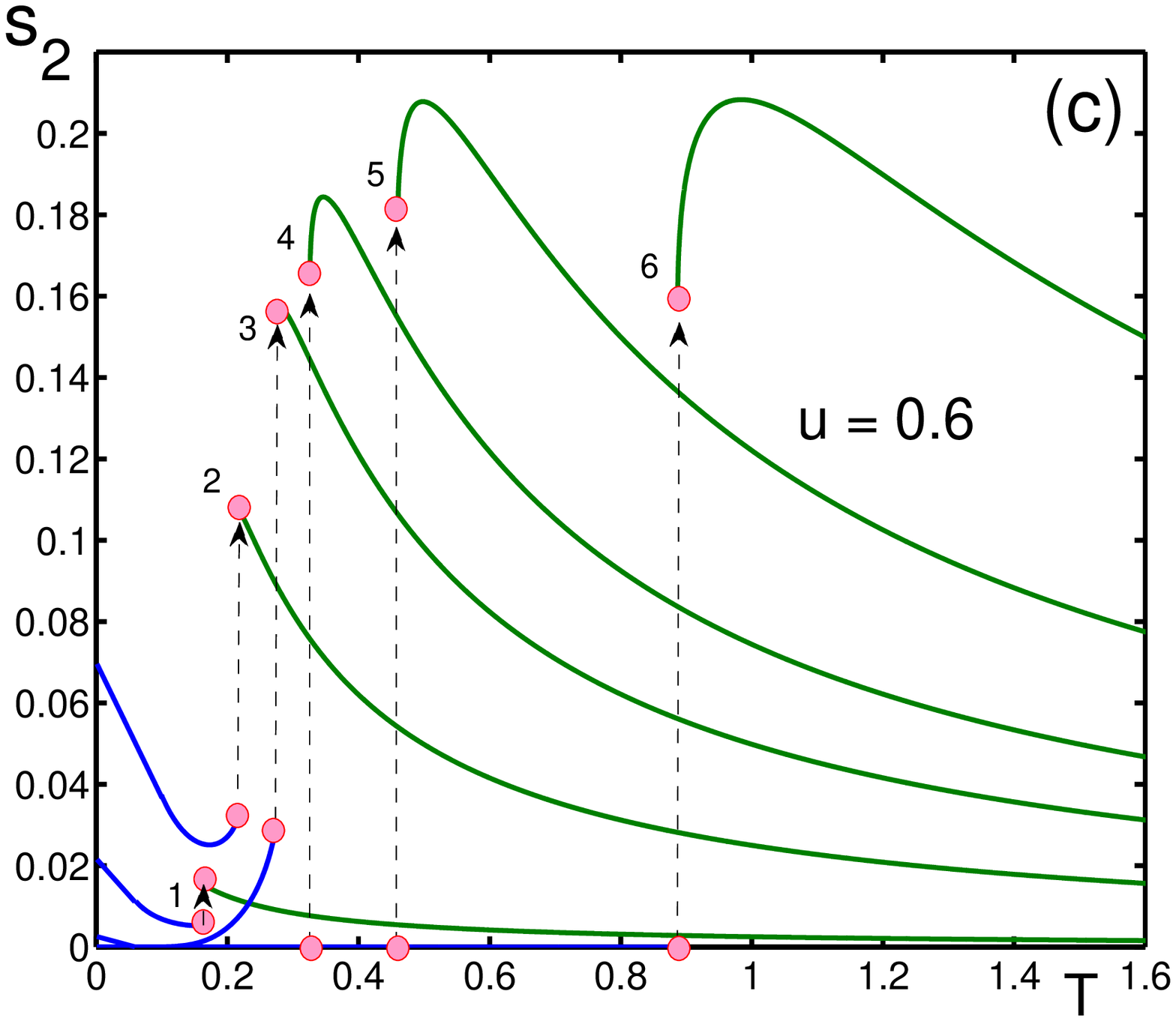}  } }
\caption{Macromolecule order parameters: (a) $w$; (b) $s_1$; (c) $s_2$. 
Dependence on dimensionless temperature is shown for the repulsion parameter 
$u = 0.6$ and varying strength of external field: (1) $h = 0.01$; (2) $h = 0.1$;
(3) $h = 0.2$; (4) $h = 0.3$; (5) $h = 0.5$; (6) $h = 1$. The corresponding 
transition temperatures are: (1) $T_0= 0.164$; (2) $T_0 =0.217$; 
(3) $T_0 = 0.276$; (4) $T_0 = 0.326$; (5) $T_0 = 0.460$; (6) $T_0 = 0.877$.}
\label{fig:Fig.2}
\end{figure}

\begin{figure}[ht]
\vspace{9pt}
\centerline{
\hbox{ \includegraphics[width=8.5cm]{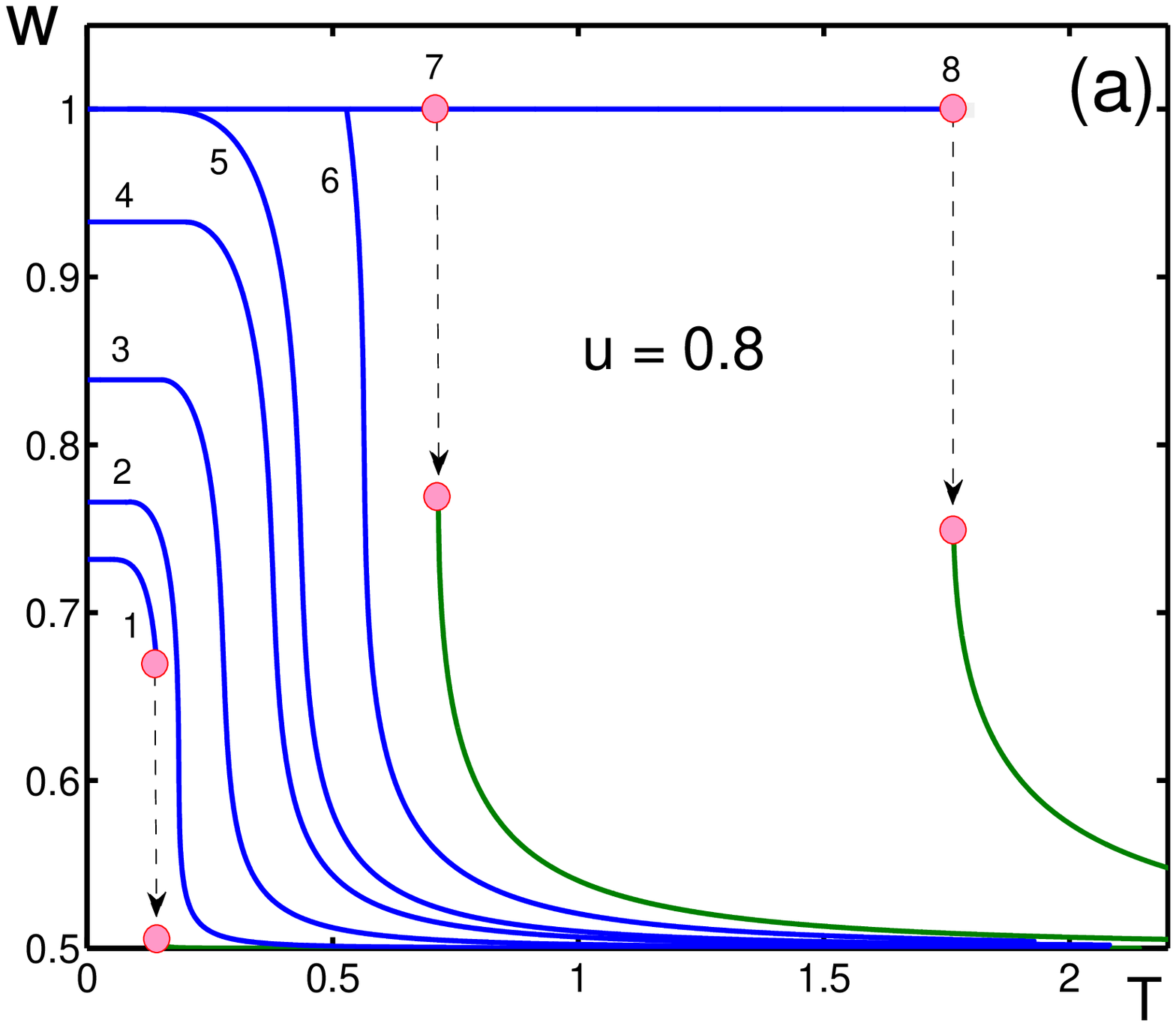} \hspace{2cm}
\includegraphics[width=8.5cm]{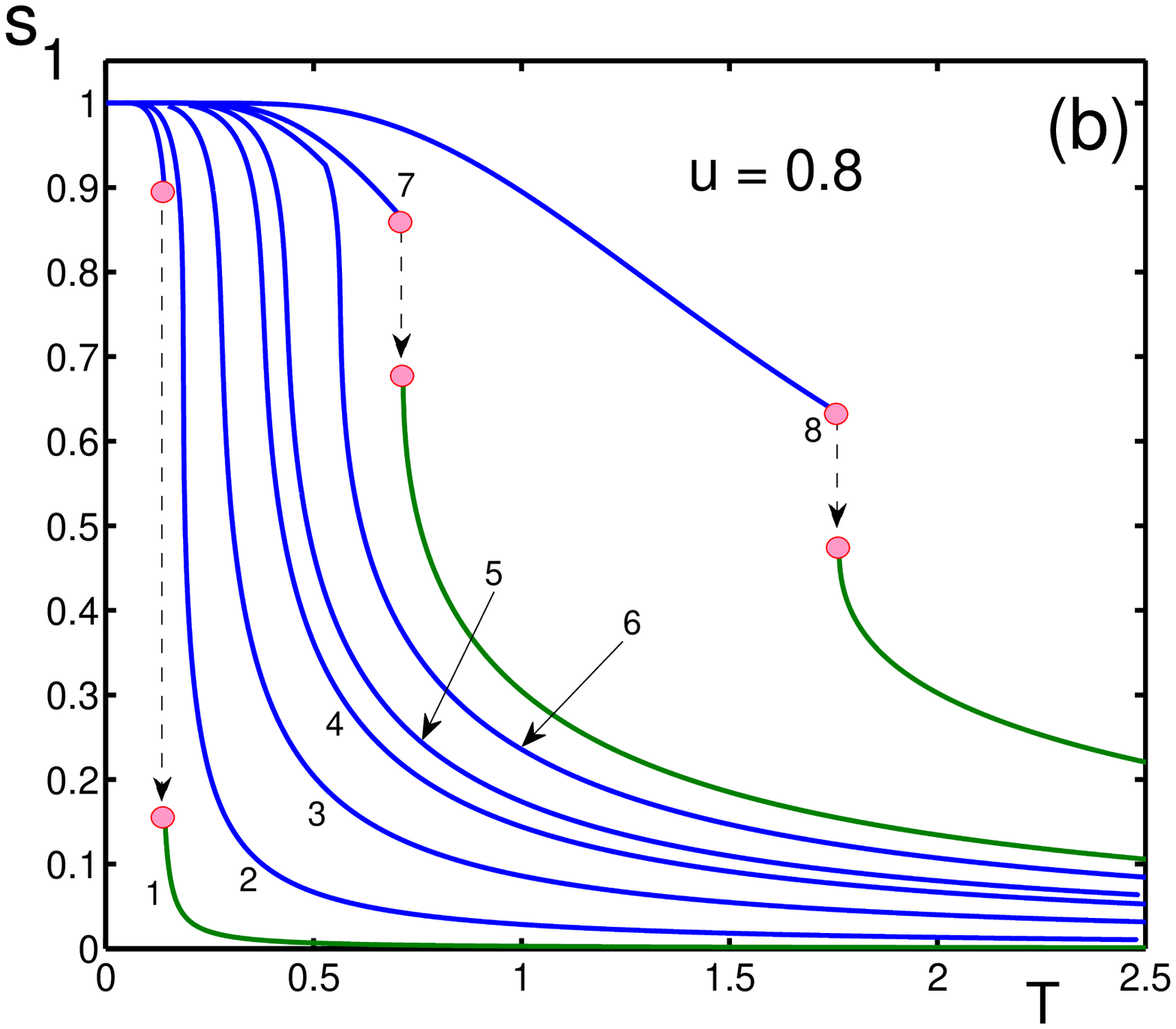} } }
\vspace{9pt}
\centerline{
\hbox{ \includegraphics[width=8.5cm]{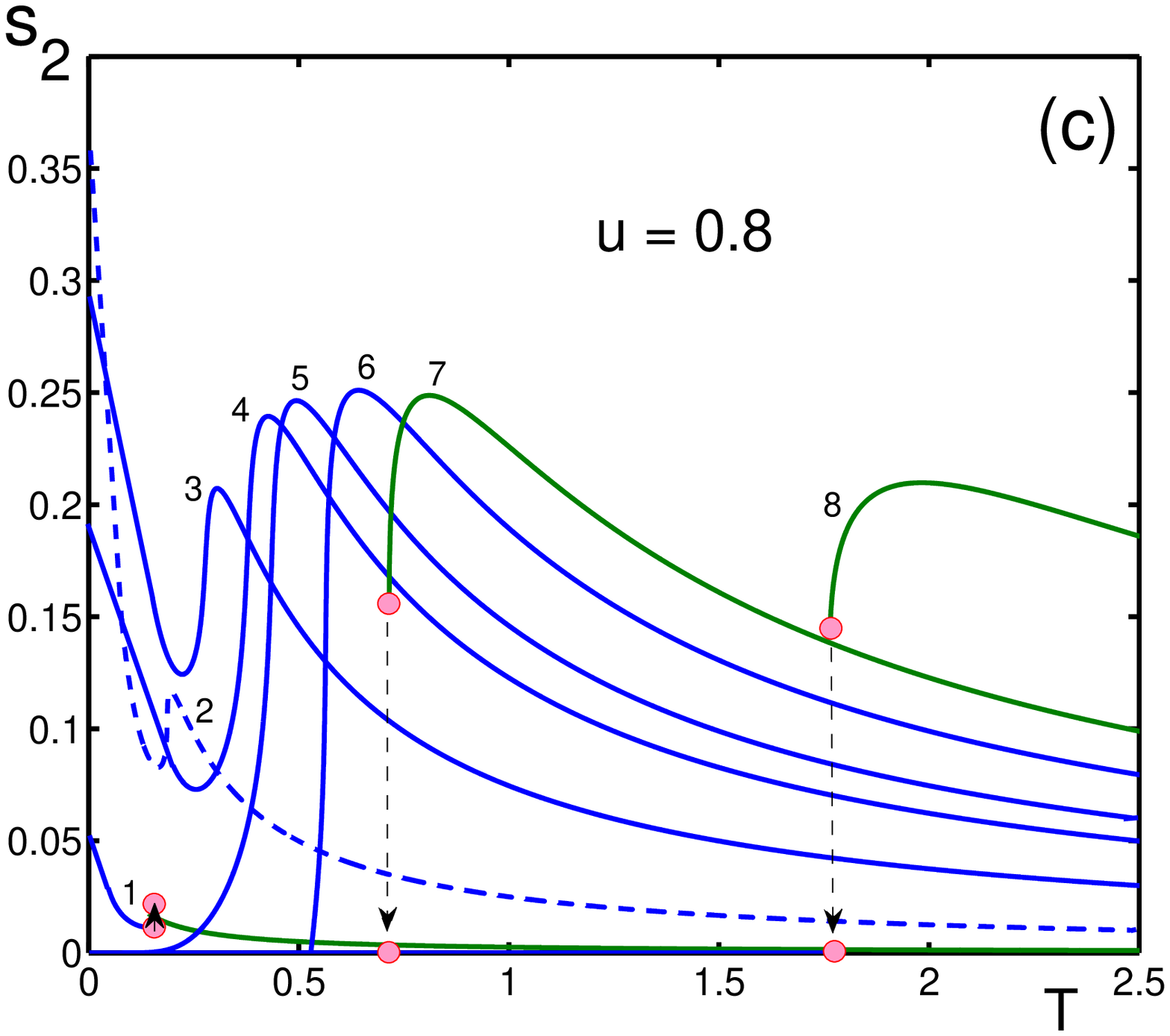}  } }
\caption{Macromolecule order parameters: (a) $w$; (b) $s_1$; (c) $s_2$. 
Dependence on dimensionless temperature is shown for $u = 0.8$ and varying 
external field: (1) $h = 0.01$; (2) $h = 0.1$; (3) $h = 0.2$; (4) $h = 0.3$; 
(5) $h = 0.5$; (6) $h = 0.8$; (7) $h = 1$; (8) $h = 2$. The first-order 
transition temperatures are: (1) $T_0 = 0.143$; (7) $T_0 = 0.715$; 
(8) $T_0 = 1.764$. For the values $0.01 < h < 1$, the transition is 
a sharp crossover.}
\label{fig:Fig.3}
\end{figure}

\begin{figure}[ht]
\vspace{9pt}
\centerline{
\hbox{ \includegraphics[width=8.5cm]{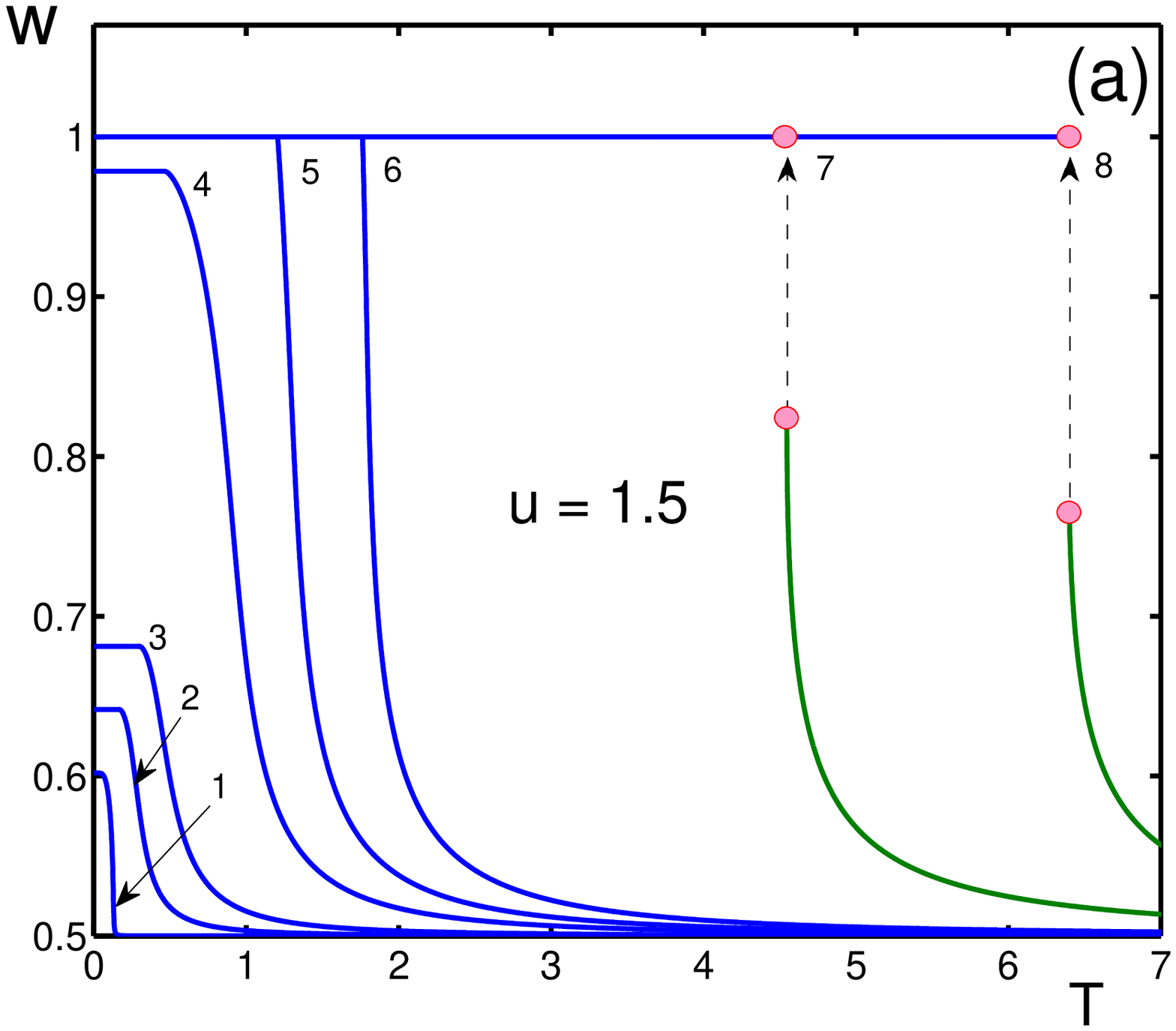} \hspace{2cm}
\includegraphics[width=8.5cm]{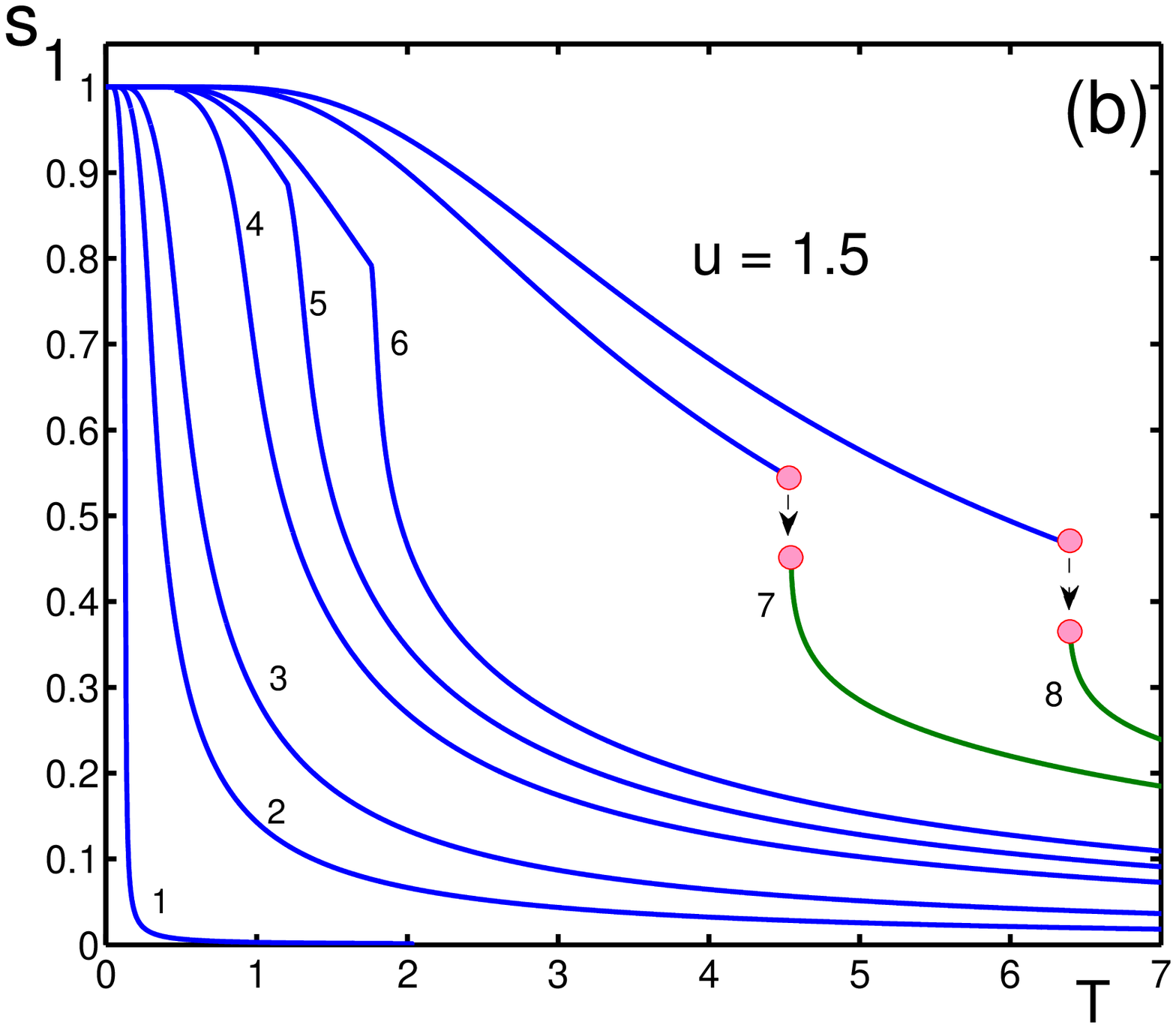} } }
\vspace{9pt}
\centerline{
\hbox{ \includegraphics[width=8.5cm]{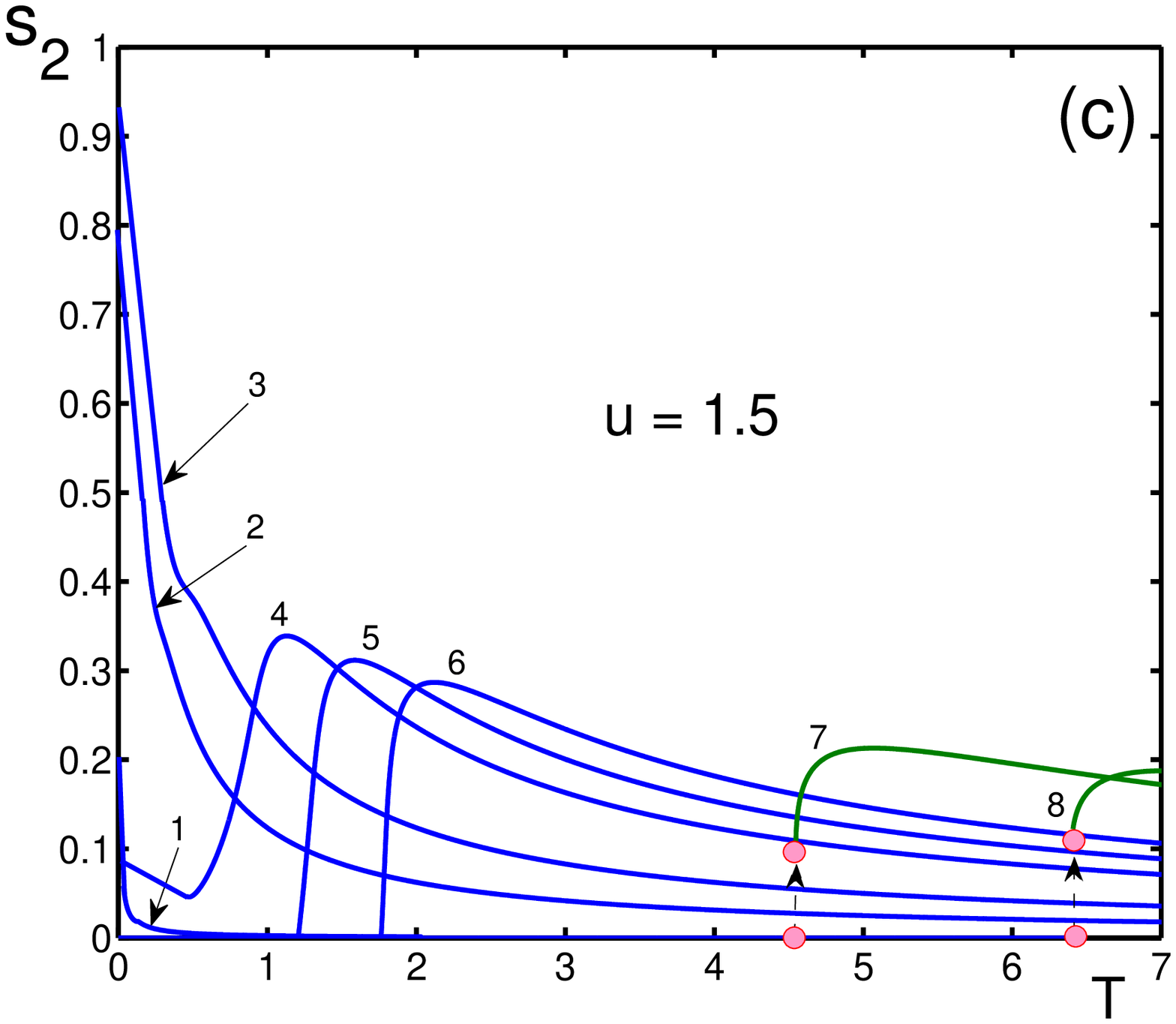}  } }
\caption{Macromolecule order parameters: (a) $w$; (b) $s_1$; (c) $s_2$. 
Dependence on dimensionless temperature is shown for $u = 1.5$ and varying 
external field: (1) $h = 0.01$; (2) $h = 0.5$; (3) $h = 1$; (4) $h = 2$; 
(5) $h = 2.5$; (6) $h = 3$; (7) $h = 5$; (8) $h = 6$. The transition 
temperatures are: (7) $T_0 = 4.547$; (8) $T_0 = 6.4$. For the values 
$h < 5$, the transition occurs as a sharp crossover.}
\label{fig:Fig.4}
\end{figure}

\begin{figure}[ht]
\vspace{9pt}
\centerline{
\hbox{ \includegraphics[width=9cm]{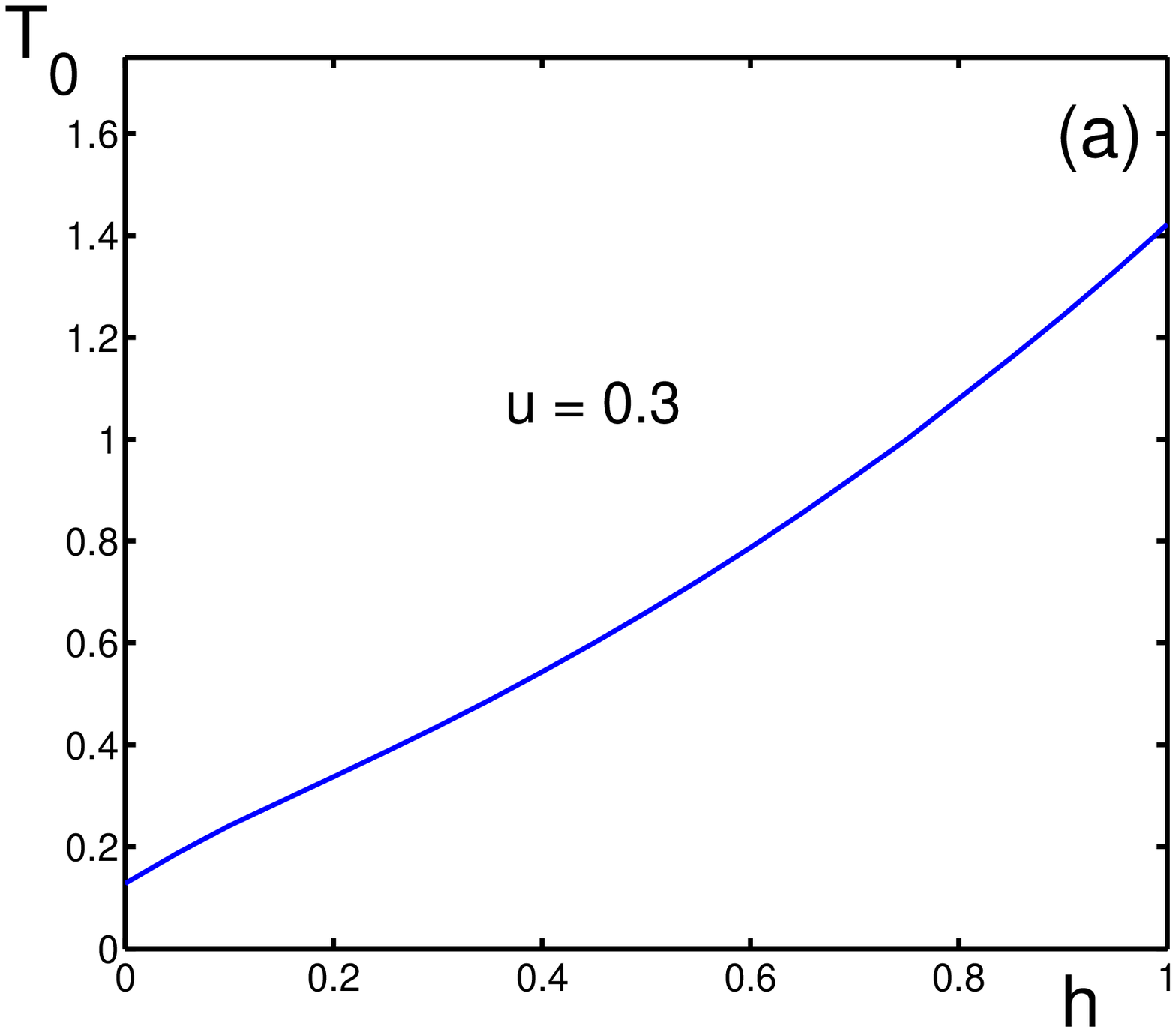} \hspace{2cm}
\includegraphics[width=9cm]{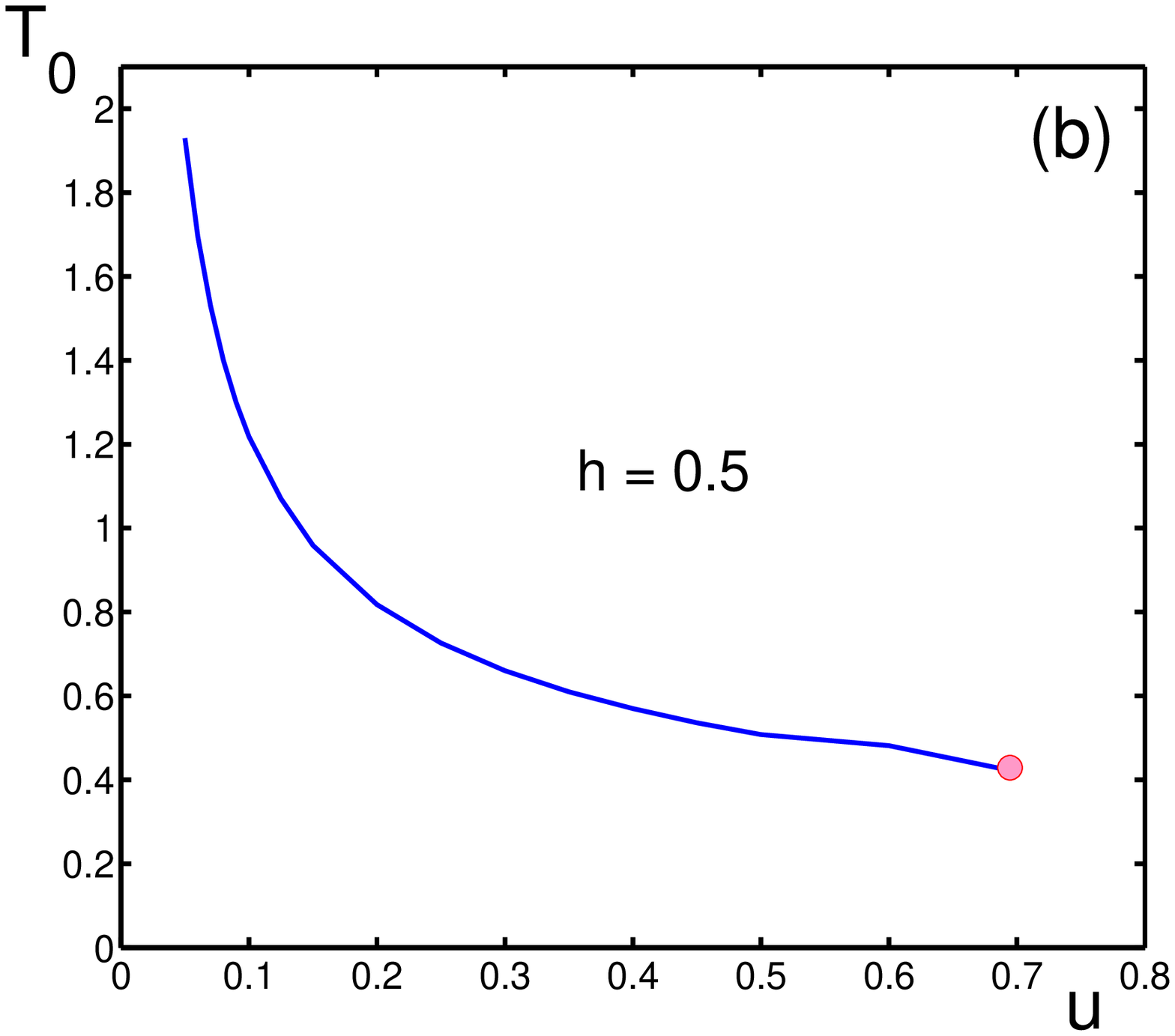} } }
\caption{First-order transition temperature $T_0$: (a) as a function of 
$h$ at fixed $u = 0.3$; (b) as a function of $u$ at fixed $h = 0.5$. 
}
\label{fig:Fig.5}
\end{figure}

\vskip 4cm

\begin{figure}[ht]
\vspace{9pt}
\centerline{
\hbox{ \includegraphics[width=10cm]{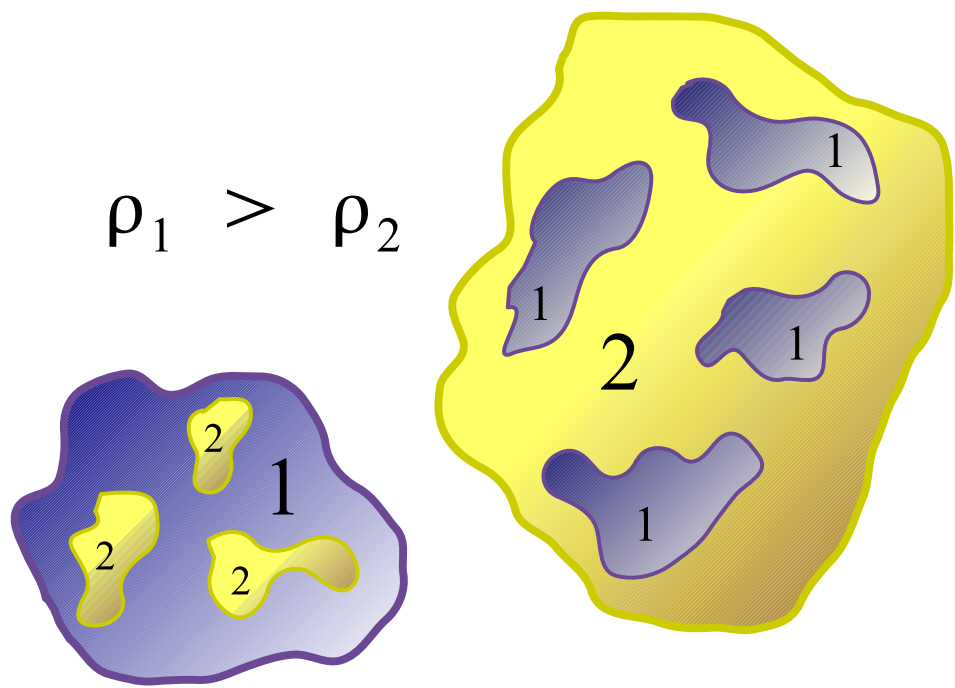}  } }
\caption{Qualitative illustration of the macromolecule swelling 
provoked by multiscale mesoscopic density fluctuations.
}
\label{fig:Fig.6}
\end{figure}

\end{document}